\documentclass[12pt]{article}
\usepackage{graphicx}
\begin{document}


\def\a{\alpha}
\def\b{\beta}
\def\c{\varepsilon}
\def\d{\delta}
\def\e{\epsilon}
\def\f{\phi}
\def\g{\gamma}
\def\h{\theta}
\def\k{\kappa}
\def\l{\lambda}
\def\m{\mu}
\def\n{\nu}
\def\p{\psi}
\def\q{\partial}
\def\r{\rho}
\def\s{\sigma}
\def\t{\tau}
\def\u{\upsilon}
\def\v{\B}
\def\w{\omega}
\def\x{\xi}
\def\y{\eta}
\def\z{\zeta}
\def\D{\Delta}
\def\G{\Gamma}
\def\H{\Theta}
\def\L{\Lambda}
\def\F{\Phi}
\def\P{\Psi}
\def\S{\Sigma}

\def\o{\over}
\newcommand{\gsim}{ \mathop{}_{\textstyle \sim}^{\textstyle >} }
\newcommand{\lsim}{ \mathop{}_{\textstyle \sim}^{\textstyle <} }
\newcommand{\vev}[1]{ \left\langle {#1} \right\rangle }
\newcommand{\bra}[1]{ \langle {#1} | }
\newcommand{\ket}[1]{ | {#1} \rangle }
\newcommand{\EV}{ {\rm eV} }
\newcommand{\KEV}{ {\rm keV} }
\newcommand{\MEV}{ {\rm MeV} }
\newcommand{\GEV}{ {\rm GeV} }
\newcommand{\TEV}{ {\rm TeV} }
\def\diag{\mathop{\rm diag}\nolimits}
\def\Spin{\mathop{\rm Spin}}
\def\SO{\mathop{\rm SO}}
\def\O{\mathop{\rm O}}
\def\SU{\mathop{\rm SU}}
\def\U{\mathop{\rm U}}
\def\Sp{\mathop{\rm Sp}}
\def\SL{\mathop{\rm SL}}
\def\tr{\mathop{\rm tr}}

\def\IJMP{Int.~J.~Mod.~Phys. }
\def\MPL{Mod.~Phys.~Lett. }
\def\NP{Nucl.~Phys. }
\def\PL{Phys.~Lett. }
\def\PR{Phys.~Rev. }
\def\PRL{Phys.~Rev.~Lett. }
\def\PTP{Prog.~Theor.~Phys. }
\def\ZP{Z.~Phys. }
\newcommand{\bear}{\begin{array}}  \newcommand{\eear}{\end{array}}
\newcommand{\bea}{\begin{eqnarray}}  \newcommand{\eea}{\end{eqnarray}}
\newcommand{\beq}{\begin{equation}}  \newcommand{\eeq}{\end{equation}}
\newcommand{\bef}{\begin{figure}}  \newcommand{\eef}{\end{figure}}
\newcommand{\bec}{\begin{center}}  \newcommand{\eec}{\end{center}}
\newcommand{\non}{\nonumber}  \newcommand{\eqn}[1]{\beq {#1}\eeq}
\newcommand{\lmk}{\left(}  \newcommand{\rmk}{\right)}
\newcommand{\lkk}{\left[}  \newcommand{\rkk}{\right]}
\newcommand{\lhk}{\left \{ }  \newcommand{\rhk}{\right \} }
\newcommand{\del}{\partial}  \newcommand{\abs}[1]{\vert{#1}\vert}
\newcommand{\vect}[1]{\mbox{\boldmath${#1}$}}
\newcommand{\bib}{\bibitem} \newcommand{\new}{\newblock}
\newcommand{\la}{\left\langle} \newcommand{\ra}{\right\rangle}
\newcommand{\bfx}{{\bf x}} \newcommand{\bfk}{{\bf k}}
\newcommand{\gtilde} {~ \raisebox{-1ex}{$\stackrel{\textstyle >}{\sim}$} ~} 
\newcommand{\ltilde} {~ \raisebox{-1ex}{$\stackrel{\textstyle <}{\sim}$} ~}
\newcommand{\gtrsim}{ \mathop{}_{\textstyle \sim}^{\textstyle >} }
\newcommand{\lesssim}{ \mathop{}_{\textstyle \sim}^{\textstyle <} }
\newcommand{\ds}{\displaystyle}
\newcommand{\bi}{\bibitem}
\newcommand{\lar}{\leftarrow}
\newcommand{\rar}{\rightarrow}
\newcommand{\lrar}{\leftrightarrow}
\def\Frac#1#2{{\displaystyle\frac{#1}{#2}}}
\def\labelenumi{(\roman{enumi})}
\def\SEC#1{Sec.~\ref{#1}}
\def\FIG#1{Fig.~\ref{#1}}
\def\EQ#1{Eq.~(\ref{#1})}
\def\EQS#1{Eqs.~(\ref{#1})}
\def\lrf#1#2{ \left(\frac{#1}{#2}\right)}
\def\lrfp#1#2#3{ \left(\frac{#1}{#2}\right)^{#3}}

\newcommand{\fa}{F_{\cal A}}
\newcommand{\A}{{\cal A}}
\newcommand{\B}{{\cal B}}
\newcommand{\gef}{{\cal G}^{\rm (eff)}_\Phi}
\newcommand{\gep}{{\cal G}^{\rm (eff)}_\Psi}

\baselineskip 0.7cm

\begin{titlepage}

\begin{flushright}
\hfill DESY 06-072\\
\hfill hep-ph/0605297\\
\hfill August, 2006\\
\end{flushright}

\vskip 1.35cm
\begin{center}
{\large \bf
The Gravitino-Overproduction Problem in Inflationary Universe}
\vskip 1.2cm
Masahiro Kawasaki$^{1}$, Fuminobu Takahashi$^{1,2}$ and T. T. Yanagida$^{3,4}$
\vskip 0.4cm

${}^1${\it Institute for Cosmic Ray Research,
     University of Tokyo, \\Chiba 277-8582, Japan}\\
${}^2${\it Deutsches Elektronen Synchrotron DESY, Notkestrasse 85,\\
22607 Hamburg, Germany}\\
${}^3${\it Department of Physics, University of Tokyo,\\
     Tokyo 113-0033, Japan}\\
 ${}^4${\it Research Center for the Early Universe, University of Tokyo,\\
     Tokyo 113-0033, Japan}

\vskip 1.5cm

\abstract{ We show that the gravitino-overproduction problem is
prevalent among inflation models in supergravity. An inflaton field
$\phi$ generically acquires (effective) non-vanishing auxiliary field
$\gef$, if the K\"ahler potential is non-minimal. The inflaton field
then decays into a pair of the gravitinos. We extensively study the
cosmological constraints on $\gef$ for a wide range of the gravitino
mass.  For many inflation models we explicitly estimate $\gef$, and
show that the gravitino-overproduction problem severely constrains the
inflation models, unless such an interaction as $K = \kappa/2
\,|\phi|^2 z^2 + {\rm h.c.}$ is suppressed (here $z$ is the field
responsible for the supersymmetry breaking). We find that many of them
are already excluded or on the verge of, if $\kappa \sim O(1)$.  }
\end{center}
\end{titlepage}

\setcounter{page}{2}

\section{Introduction}
\label{sec:1}

The gravitino is the most important prediction of unified theory of
quantum mechanics and general relativity such as the superstring
theory (i.e. supergravity (SUGRA) at low energies)~\cite{superstring}.
However, the presence of the gravitino leads to serious cosmological
problems depending on its mass and nature. If the gravitino is
unstable and has a mass $m_{3/2}$ in the range from $O(100)$ GeV to
$O(10)$ TeV, the decay of the gravitino destroys light elements
produced by the big-bang nucleosynthesis (BBN). To keep the success of
BBN the reheating temperature $T_R$ after inflation should be lower
than $O(10^{6-8})$ GeV suppressing the gravitino production by thermal
scattering. On the other hand, if the gravitino is light as $m_{3/2} <
O(10)$ GeV and it is stable (that is, the lightest supersymmetric
particle (LSP)), the reheating temperature should satisfy $T_R
\lesssim O(10^{7}){\rm\, GeV}(m_{3/2}/1 {\rm\,GeV})$ for $m_{3/2}
\gtrsim 100$ keV for the gravitino density not to exceed the observed
dark matter density.

In a recent article~\cite{KTY-I}, we have pointed out that there is a
new gravitino problem beside due to the thermal production of the
gravitino.  That is, an inflaton field $\phi$ has nonvanishing
supersymmetry(SUSY)-breaking auxiliary field $G_\phi$ (or more
precisely $\gef$ as will be defined later) in most of inflation models
in SUGRA, which gives rise to an enhanced decay of the inflaton into a
pair of gravitinos, if the K\"ahler potential is non-minimal.  Thus,
we have stringent constraints on the (effective) auxiliary field
$\gef$ to suppress the production of gravitinos in the inflaton
decay~\cite{KTY-I}. This gravitino production in inflaton decay is
more effective for lower reheating temperature, while the production
by particle scatterings in the thermal bath is more important for
higher temperature $T_R$. Therefore, the direct gravitino production
discussed in this paper is complementary to the thermal gravitino
production, and the former may put severe constraints on inflation
models together with the latter.

The purpose of this paper is to discuss this new gravitino problem in
a broad mass range of the gravitino including $m_{3/2} \simeq
O(100)$TeV region suggested from anomaly-mediated SUSY breaking
models~\cite{Randall:1998uk}.  We assume, in the present analysis,
that there is no entropy production after the end of reheating by the
inflaton decay.  However, we briefly discuss, in the last section of
this paper, the case that a late-time entropy production takes place.

In Sec.~\ref{sec:2} we briefly review the gravitino problem in
cosmology and in Sec.~\ref{sec:3} we calculate the abundance of
gravitinos produced by particle scatterings in the thermal bath and
show cosmological constraints on the reheating temperature $T_R$. In
Sec.~\ref{sec:4} we discuss the enhanced decay of the inflaton into a
pair of gravitinos and give cosmological constraints on the
(effective) auxiliary field $\gef$. In Sec.~\ref{sec:5} we explicitly
calculate the precise value of $\gef$ for inflation models in SUGRA to
demonstrate how severe the new constraints are. The last section is
devoted to conclusions.

\section{Gravitino problem}
\label{sec:2}

The gravitino is the SUSY partner of the graviton in SUGRA and it
acquires a mass in a range of $ {\cal O}(100)\ {\rm GeV}- {\cal
O}(10)$ TeV in gravity-mediated SUSY-breaking models~\footnote{
Although the gravitino mass can be either much
lighter~\cite{Ellis:1984kd} or much heavier~\cite{Ellis:1984bs} in
no-scale models, we do not consider such possibilities in this paper.
}.  Such a gravitino is likely unstable and its lifetime is very long
because interactions of the gravitino are suppressed by inverse powers
of the reduced Planck scale $M_P$.  The gravitino dominantly decays
into the standard-model (SM) particles and their superpartners, which
may produce a large entropy and destroy the light elements synthesized
in BBN.  As a result, the predictions of BBN may be significantly
changed unless the primordial abundance of the gravitino is
sufficiently small \cite{Weinberg:zq}.

In gauge-mediated SUSY-breaking models~\cite{GMSB}, the gravitino is
light ($m_{3/2} \lesssim 10$~GeV) and stable. In this case the
gravitino may give too much contribution to the present cosmic density
of the universe.

In the inflationary universe, the primordial gravitino is once diluted
but it is produced during reheating epoch after the inflation.  Thus,
even in the inflationary models, we may still have the gravitino
problem~\cite{Krauss:1983ik}.  As shown in the next section, this
leads to very stringent constraints on the reheating temperature $T_R$
since the gravitino abundance is approximately proportional to
$T_R$. The constraints are given in \cite{BBNwX_OLD,Kawasaki:1994af,
Protheroe:dt, Holtmann:1998gd,Jedamzik:1999di,Kawasaki:2000qr,
Kohri:2001jx,Cyburt:2002uv} for the unstable gravitino and in
\cite{Moroi:1993mb} for the stable one.

\section{Thermal production of gravitinos and cosmological constraints
on the reheating temperature $T_R$}
\label{sec:3}

In this section we show the abundance of the gravitinos thermally
produced after inflation and derive constraints on the reheating
temperature.

During reheating the gravitino is produced through scatterings of
particles in thermal bath. The interactions of the gravitino with a
gauge multiplet ($A_{\mu},\lambda$) and a chiral multiplet
($\eta,\chi$) are described by
\begin{eqnarray}
    {\cal L} &=&
    -\frac{1}{\sqrt{2}M_P} D_{\nu} \eta^{\dagger}
    \bar{\psi}_\mu \gamma^\nu \gamma^\mu \chi_R
    -\frac{1}{\sqrt{2}M_P} D_{\nu} \eta
    \bar{\chi}_L \gamma^\mu \gamma^\nu \psi_\mu
    \nonumber \\ &&
    -\frac{i}{8M_P} \bar{\psi}_\mu
    \left[ \gamma^\nu , \gamma^\rho \right] \gamma^\mu
    \lambda F_{\nu\rho},
\end{eqnarray}
where $F_{\nu\rho}$ is the field strength of the gauge field.  (Here,
$D_\nu$ denotes the covariant derivative and $\chi_R$ satisfies
$(1-\gamma_5)\chi_R=0$.)  The thermally averaged cross section of the
gravitino production for an $SU(N)$ super Yang-Mills model with
$n_{\rm f}$ pairs of fundamental and anti-fundamental chiral
superfields is calculated in Ref.~\cite{Bolz:2000fu} as
\begin{eqnarray}
    \langle \sigma v_{\rm rel} \rangle &=& 
    \left[ 1 + \left( \frac{m_{\tilde{g}}^2}{3m_{3/2}^2} \right) 
    \right]
    \frac{3 g^2 (N^2-1)}{32\pi M_P^2}
    \nonumber \\ &&
    \times \frac{\pi^2}{\zeta(3)}
    \left\{
        \left[ \ln (T^2/m_{g,{\rm th}}^2) + 0.3224 \right]
        ( N + n_{\rm f} )
        + 0.5781 n_{\rm f}
    \right\},
\end{eqnarray}
where $m_{\tilde{g}}$ is the gaugino mass and $m_{g,{\rm th}}$ is the
thermal mass of the gauge boson which is given as $m_{g,{\rm th}}^2 =
(1/6) g^2 ( N + n_{\rm f} ) T^2$.

Solving the Boltzmann equation with the above cross section, one can
obtain the gravitino-to-entropy ratio $Y_{3/2}$ which is well
approximated by~\cite{Kawasaki:2004yh}
\begin{eqnarray}
    \label{eq:Yx-new}
    Y_{3/2} &\simeq& 
    1.9 \times 10^{-12}\left[ 1+ 
    \left(\frac{m_{\tilde{g}_3}^2}{3m_{3/2}^2}\right)\right]
    \left( \frac{T_{\rm R}}{10^{10}\ {\rm GeV}} \right)
    \nonumber \\ 
    & \times & 
    \left[ 1 
        + 0.045 \ln \left( \frac{T_{\rm R}}{10^{10}\ {\rm GeV}} 
        \right) \right]
    \left[ 1 
        - 0.028 \ln \left( \frac{T_{\rm R}}{10^{10}\ {\rm GeV}} ,
        \right) \right],
\end{eqnarray}
where we have taken $N=3$ for QCD and $m_{\tilde{g}_3}$ is the gluino
mass evaluated at $T=T_R$.  Notice that the gravitino abundance is
roughly proportional to $T_R$.

For the gravitino of a relatively large mass $ \gtrsim 100$~GeV, it
likely decays to the SM particles and their superpartners.  In that
case, high energy photons and hadrons emitted in the gravitino decay
may destroy the light elements (D,$^3$He, $^4$He, $^7$Li, $\cdots$)
and hence spoil the success of BBN.  Since the gravitino abundance is
approximately proportional to $T_R$, we obtain an upper bound on the
reheating temperature after inflation.

Energetic photons from the radiative decay of gravitino ($\psi_{\mu}
\rightarrow \gamma + \tilde{\gamma}$) deconstruct D, which gives an
upper bound $T_R \lesssim 10^{9}$~GeV for $m_{3/2} \simeq 1 -
3$~TeV. They also cause an overcreation of $^3$He due to
photo-dissociation of $^4$He, which leads to the most stringent
constraint on $T_R$ as $T_R \simeq 10^{6}-10^{9}$~GeV for $m_{3/2}
\simeq 100$~GeV $- 1$~TeV.

However, it was found in Ref.~\cite{Kawasaki:2004yh} that the hadronic
decay gives a more stringent constraint on the abundance of gravitinos
and equivalently on the reheating temperature $T_R$ because mesons and
nucleons produced in the decay and subsequent hadronization processes
significantly affect BBN. In particular, when the branching ratio into
hadrons is $\sim 1$ as expected for the gravitino decaying into gluino
and gluon, the effect of the hadronic decay is much more serious than
the radiative one.

In the case of $B_h=1$ ($B_h$: the branching ratio of the hadronic
decay), the upper bound on $T_R$ for relatively light gravitino
$m_{3/2}\simeq 0.1- 0.2$~TeV comes from the overproduction of $^3$He
as
\begin{equation}
\label{eq:rtemp_0}
   T_R ~\lesssim ~(1-4)\times 10^6~{\rm GeV}
   ~~~~~~~ {\rm for} ~~ m_{3/2} \simeq 0.1 - 0.2~{\rm TeV}
   ~~~~(B_h \simeq 1),
\end{equation}
Here we conservatively assume $m_{\tilde{g}} \ll m_{3/2}$.  For
$m_{3/2} \simeq 0.2-1$~TeV, non-thermal production of $^6$Li sets the
very stringent constraint as,
\begin{equation}
   \label{eq:rtemp_1}
   T_R ~ \lesssim ~ 3 \times 10^{5} -  4 \times 10^{6}~{\rm GeV}
   ~~~~ {\rm for} ~~ m_{3/2} \simeq 0.2 - 2~{\rm TeV}
   ~~~~(B_h \simeq 1).
\end{equation}
For larger gravitino mass the destruction of D gives the stringent
constraint,
\begin{equation}
   \label{eq:rtemp_2}
    T_R ~\lesssim ~  5\times 10^{5} -  1\times 10^{8}~{\rm GeV}
   ~~~~ {\rm for} ~~ m_{3/2} \simeq 2 -  10~{\rm TeV}
   ~~~~(B_h \simeq 1).
\end{equation}
Since the gravitino of mass larger than 10~TeV decays before the light
elements are synthesized, the stringent constraint is not obtained
from hadro-dissociation processes.  However, the mesons (mainly pions)
produced at $\sim 1~\sec$ alter the proton-neutron ratio and increase
the abundance of $^4$He, from which the upper bound on $T_R$ is
obtained as
\begin{equation}
    T_R ~\lesssim ~(3 -10)\times 10^9~{\rm GeV}
   ~~~~ {\rm for} ~~ m_{3/2} \simeq 10 -  30~{\rm TeV}.
   ~~~~(B_h \simeq 1)
\end{equation}
For $m_{3/2} \gtrsim 30$~TeV the gravitino decay little affects BBN in
the case of $B_h =1$.

When the main decay mode is not hadronic, the above constraints become
milder.  However, even if the gravitino dominantly decays into a
photon and a photino, the hadronic branching ratio is non-vanishing
since the quark-anti-quark pair can be attached at the end of the
virtual photon line.  In this case, $B_h$ is expected to be of order
${\cal O}(\alpha_{\rm em}/4\pi) \simeq 10^{-3}$. Even such small $B_h$
makes the constraint severer than that for pure radiative decay ($B_h
=0$) as 
\begin{eqnarray}
\label{eq:unstable-g1}
    T_R  ~ \lesssim ~  1\times 10^{6}- 3\times 10^{8}~{\rm GeV}
   ~~~~ &{\rm for}& ~~ m_{3/2} \simeq 0.1 - 1~{\rm TeV} \\
\label{eq:unstable-g2}   
    T_R   ~\lesssim ~  1\times 10^{8} -  3\times  10^{8}~{\rm GeV}
   ~~~~ &{\rm for}& ~~ m_{3/2} \simeq 1 - 3~{\rm TeV} \\
\label{eq:unstable-g3}   
    T_R   ~\lesssim ~  2\times 10^{8} -  1\times  10^{9}~{\rm GeV}
   ~~~~ &{\rm for}& ~~ m_{3/2} \simeq 3 - 10~{\rm TeV} \\
   & & ~~~~~~~~~~~~~~~~~~~~~~(B_h \simeq 10^{-3}) \nonumber
\end{eqnarray}
where the upper limits on $T_R$ are imposed by $^3$He overproduction,
$^6$Li overproduction and D destruction, respectively. In the case of
$B_h =10^{-3}$, no sensible BBN bound exists for $m_{3/2} \gtrsim
10$~TeV.

The corresponding constraints on $Y_{3/2}$ which will be used later are
obtained by substituting the upper bounds on $T_R$ into (\ref{eq:Yx-new}),
\begin{eqnarray}
\label{eq:unstable-Y1}
   Y_{3/2}  & ~ \lesssim & \left\{\begin{array}{lcl}
   ~1\times 10^{-16} - 6\times 10^{-16}
   &{\rm for}    &  m_{3/2} \simeq 0.1 - 0.2~{\rm TeV} \\[0.8em]
   ~4\times 10^{-17} - 6\times 10^{-16}
   &{\rm for}    &  m_{3/2} \simeq 0.2 - 2~{\rm TeV} \\[0.8em]
   ~ 7 \times 10^{-17} - 2\times 10^{-14} 
   &{\rm for}    & m_{3/2} \simeq 2 - 10~{\rm TeV} \\[0.8em]
   ~ 6\times 10^{-13} - 2\times 10^{-12} 
   &{\rm for}    & m_{3/2} \simeq 10 - 30~{\rm TeV} \end{array}\right.
   ~~(B_h \simeq 1),\\[1em]
\label{eq:unstable-Y2}
   Y_{3/2}  &~ \lesssim & \left\{\begin{array}{lcl}
   ~1\times 10^{-16} - 5\times 10^{-14}
   &{\rm for}&  m_{3/2} \simeq 0.1 - 1~{\rm TeV} \\[0.8em]
   ~ 2\times 10^{-14} - 5\times 10^{-14}
   &{\rm for}&  m_{3/2} \simeq 1 - 3~{\rm TeV} \\ [0.8em]
    ~ 3\times 10^{-14} - 2\times 10^{-13}
   &{\rm for}&  m_{3/2} \simeq 3 - 10~{\rm TeV} \end{array}\right.
   ~~(B_h \simeq 10^{-3}). 
\end{eqnarray}

For the heavy gravitino of mass $\gtrsim 30(10)$ TeV with $B_h = 1 (10^{-3})$,
no stringent constraints are obtained from BBN.  However,
another constraint comes from the abundance of the LSP produced by the
gravitino decay.  Since the gravitino decay temperature is rather low,
one LSP remains as a result of the decay of one gravitino. The relic
LSP density is
\beq 
\Omega_{\rm LSP} h^2 \simeq 0.052 \lrf{m_{\rm
LSP}}{100{\rm\,GeV}} \lrf{T_R}{10^{10}{\rm\,GeV}}, 
\eeq
where $m_{\rm LSP}$ is the LSP mass, and we have conservatively
neglected the contribution from the thermally produced LSPs. According
to the recent WMAP result~\cite{Spergel:2006hy}, the dark matter
density is $\Omega_{\rm DM} h^2 \simeq 0.11 \pm 0.01$ ($h$: Hubble
parameter in units of 100km/s/Mpc). Requiring the LSP density smaller
than the upper bound on the dark matter density at 95 \% C.L., we
obtain
\beq
T_R \;\lesssim\; 2.5 \times 10^{10} \lrfp{m_{\rm LSP}}{100{\rm\,GeV}}{-1} {\rm GeV},
\label{eq:const-from-lsp}
\eeq
which is applicable for the unstable gravitinos. This bound is
important especially for the gravitino heavier than $30 (10)$ TeV,
which falls in the range suggested from anomaly-mediated models of
SUSY breaking~\cite{Randall:1998uk}.  In the anomaly-mediated SUSY
breaking models, the LSP is mostly composed of the wino $\tilde{W}$
and its mass is related to the gravitino mass as
\begin{equation}
\label{eq:winomass}
    m_{\tilde{W}} \;=\; \frac{\beta_2}{g_2} \,m_{3/2} \simeq 2.7 \times 10^{-3} m_{3/2},
\end{equation}
where $\beta_2$ and $g_2$ are the beta function and the gauge coupling
of $SU(2)_L$. Since the thermal relic of the wino LSP is less than the
observed dark matter abundance as long as $m_{\tilde{W}} \lesssim
2{\rm\, TeV}$~\cite{Ibe:2004gh}, we obtain
\beq
T_R \;\lesssim\; 9.3 \times 10^{9} \lrfp{m_{3/2}}{100{\rm\,TeV}}{-1} {\rm GeV},
\label{eq:const-from-winolsp}
\eeq
for $m_{3/2} \lesssim 7 \times 10^{2}{\rm\, TeV}$ in the
anomaly-mediated SUSY breaking models with the wino LSP.

When the gravitino is light ($\lesssim 10$~GeV) which is expected in
gauge-mediated SUSY breaking models, the gravitino may be the LSP and
hence stable.  Since the cosmic density of the gravitino should be
less than the dark matter density of the universe~\cite{Moroi:1993mb},
we obtain the constraint
\begin{equation}
\label{eq:stable-g}
    T_R ~\lesssim ~3\times 10^7~{\rm GeV} 
   \lrfp{m_{\tilde{g}_3}}{500{\rm\,GeV}}{-2}   \left(\frac{m_{3/2}}{1{\rm \,GeV}}\right)
   ~~~~~{\rm for}~~~  m_{3/2}\simeq
    10^{-4} - 10~{\rm GeV},
\end{equation}
where we have omitted the logarithmic corrections.  For $1$~keV $
\lesssim m_{3/2} \lesssim 10^{-4}$~GeV, the upper-bound on $T_R$ is of
the order of $100$~GeV,
\begin{equation}
\label{eq:stable-g2}
  T_R \lesssim O(100)~{\rm GeV} ~~~~~~{\rm for}~~~ 
    m_{3/2}\simeq 1~{\rm keV} - 10^{-4}~{\rm GeV}.
\end{equation}
When the gravitino is lighter than $1$~keV, no constraint comes from
the cosmic density. However, such a light gravitino behaves as warm or
hot dark matter component and affects the power spectrum of the
density fluctuations through free streaming. This may extend the bound
(\ref{eq:stable-g2}) to $m_{3/2} \sim O(10)$~eV~\cite{Viel:2005qj}.

\section{Gravitino production in inflaton decay and its cosmological constraints}
\label{sec:4}
In this section we first estimate the decay rate of the inflaton into
a pair of the gravitinos and clarify a condition under which this
decay channel becomes effective. Then we discuss cosmological
constraints on such a decay for a broad range of the gravitino mass:
$m_{3/2} = 1{\rm\,keV}- 100$\,TeV, in order to show how severe this
gravitino-oveproduction problem is.

\subsection{Inflaton decay into a pair of gravitinos}
\label{sec:4-1}
The relevant interactions for the decay of an inflaton field $\phi$
into a pair of the gravitinos are~\cite{WessBagger}
\bea
   e^{-1}\mathcal{L} &=&
   - \frac{1}{8} \epsilon^{\mu\nu\rho\sigma}
   \left( G_\phi \partial_\rho \phi + G_z \partial_\rho z -  
{\rm h.c.}
     \right)
   \bar \psi_\mu \gamma_\nu \psi_\sigma\non\\
   &&
   - \frac{1}{8} e^{G/2} \left( G_\phi \phi + G_z z +{\rm h.c.}
    \right)
   \bar\psi_\mu \left[\gamma^\mu,\gamma^\nu\right] \psi_\nu,
   \label{eq:inf2gravitino}
\eea
where $\psi_\mu$ is the gravitino field, and we have chosen the
unitary gauge in the Einstein frame with the Planck units, $M_P =1$.
We have defined the total K\"ahler potential, $G=K+{\rm ln}\,|W|^2$,
where $K$ and $W$ are the K\"{a}hler potential and superpotential,
respectively. Here and in what follows a subscript $i$ denotes a
derivative with respect to the field $i$, while a superscript is
obtained by multiplying with $g^{i j^*}$, the inverse of the K\"ahler
metric $g_{ij^*} \equiv G_{i j^*}$.  The SUSY breaking field $z$ is
such that it sets the cosmological constant to be zero, i.e., $G^z G_z
\simeq 3$~\footnote{
Throughout this paper we assume that the $D$-term potential is negligible.
In a broad class of the SUSY breaking models,  the $z$ field may not be
the only field that has a sizable $F$-term, and $|G_z|$ could differ from
$\sqrt{3}$. However, the following arguments remain virtually intact as long as 
$|G_z| \simeq O(1)$; if $G^z G_z$ decreases by one order of magnitude,
the constraints on inflation models would become relaxed by the same amount. 
}, and we assume that $z$ is a singlet under any symmetries as in the
gravity-mediated SUSY breaking models. In fact, the existence of a
singlet (and elementary) field $z$ with a nonzero F-term of $O(m_{3/2}
M_P)$ is a generic prediction of the gravity-mediated SUSY breaking
models. This is true even in the case of the dynamical SUSY
breaking~\cite{Witten:1981nf}, because the gauginos would become much
lighter than squarks and sleptons otherwise~\cite{Banks:1993en}~\footnote{
Note, however, that it is possible, though complicated, to generate a sizable 
gaugino mass by introducing extra chiral superfields
in the adjoint representation of the gauge group, rather than a singlet~\cite{Dine:1992yw}.
}.
Later we will give a comment on the case that $z$ is charged under
some symmetry as in the gauge-mediated and anomaly-mediated SUSY
breaking models.

It has been recently argued that the modulus and inflaton decays
produce too much gravitinos through the above
interaction~\cite{Endo:2006zj, KTY-I, Asaka:2006bv}.  Taking account
of the mixing between $\phi$ and $z$, however, the effective coupling
of the inflaton with the gravitinos is modified~\cite{Dine:2006ii}.
According to the detailed calculation of Ref.~\cite{Endo:2006tf}, we
only have to replace $G_\phi$ with $\gef$ (the relation between these
two is given in the next section).  The real and imaginary components
of the inflaton field have the same decay rate at the leading
order~\cite{Endo:2006zj,Endo:2006tf}:
\beq
\label{eq:decay-rate}
 \Gamma_{3/2} \equiv  \Gamma(\phi \rightarrow 2\psi_{3/2}) \simeq
  \frac{|\gef|^2}{288\pi}  \frac{m_\phi^5}{m_{3/2}^2 M_P^2}, 
\eeq
where we have assumed that the inflaton has a supersymmetric mass much
larger than the gravitino mass: $m_\phi \gg m_{3/2}$. Thus the decay
rate is enhanced by the gravitino mass in the denominator, which comes
from the longitudinal component of the gravitino~\footnote{
The decay can also be understood in terms of the goldstinos due to the
equivalence theorem in supergravity~\cite{eq-in-SUGRA,Endo:2006tf}.
}, as emphasized in Ref.~\cite{Endo:2006zj}.

It should be noted that the above expression for the decay rate cannot
be applicable for $H>m_{3/2}$.  The decay proceeds only if the Hubble
parameter $H$ is smaller than the gravitino mass, since the chirality
flip of the gravitino forbids the decay to proceed otherwise.
Intuitively, the gravitino is effectively massless as long as
$H>m_{3/2}$.

We should clarify another important issue: what is the longitudinal
component of the gravitino (i.e. goldstino) made of ? A similar issue
was discussed in the context of the non-thermal `gravitino' production
during preheating~\cite{Kallosh:1999jj}, and it was concluded that the
inflatino, instead of the gravitino in the low energy, is actually
created~\cite{Nilles:2001ry}~\footnote{
It should be noted, however, that the inflatinos produced during
preheating may be partially converted to the gravitinos in the low
energy, since $G_\phi$ is generically nonzero in the true
minimum~\cite{KTY-I} (the inflation model adopted in
Ref.~\cite{Nilles:2001ry} has vanishing $G_\phi$).  This effect may
further constrain the inflation models.
}.  Since the inflatino decays much earlier than the BBN
epoch~\cite{Allahverdi:2000fz}, the non-thermal `gravitino' (actually,
inflatino) production turned out to be harmless.  The reason is that
the `gravitino' production occurs in a rather early stage of the
reheating just after the inflation ends, during which the energy
stored in the inflationary sector significantly contributes to the
total SUSY breaking.  In our case, however, the situation is
completely different; the decay into the gravitinos is effective,
since we consider a cosmological epoch, $H<m_{3/2}$, when the SUSY
breaking contribution of the inflaton is subdominant. Thus, the
gravitinos produced directly by the inflaton decay should coincide
with those in the low energy.

\begin{figure}[t]
\begin{center}
\includegraphics[width=7cm]{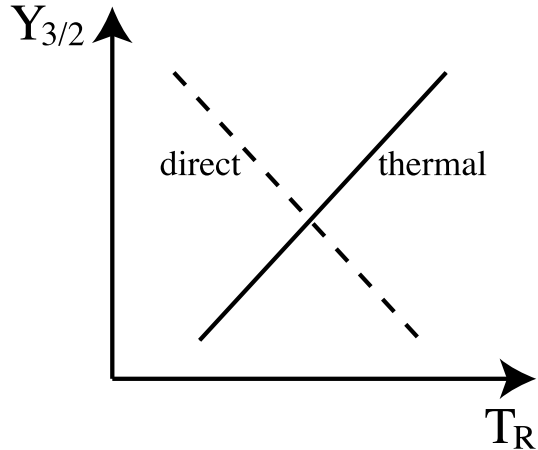}
\caption{Dependence of the gravitino-to-entropy ratio on 
the reheating temperature $T_R$. The sold line represents the
abundance of the thermally produced gravitinos, 
while the dashed line corresponds to 
that directly produced by the inflaton decay.
See  (\ref{eq:Yx-new}) and (\ref{eq:ngs}) in the text.
}
\label{fig:y32-tr}
\end{center}
\end{figure}

Let us now consider the implication of (\ref{eq:decay-rate}).  As we
will see in the next section, the effective coupling $\gef$ is
proportional to $G_\phi$, for such non-minimal interaction as
$(\kappa/2) |\phi|^2 zz +{\rm h.c.}$ in the K\"ahler potential.  The
auxiliary field $G_\phi$ represents the fractional contribution of the
inflaton to the SUSY breaking.  One might suspect that $G_\phi$ (and
therefore $\gef$) should be zero in the vacuum and such a decay does
not occur at all. However, as we will see in the next section, this is
generically not true.  To be sure, in many inflation models, the
minimum of the inflaton potential preserves SUSY, as long as the
inflaton sector is concerned. But, once we take account of the SUSY
breaking sector, the minimum slightly shifts and non-vanishing
$G_\phi$ is induced.  This means that we need to consider the scalar
potential including both the inflaton and the SUSY breaking sector
field, in order to evaluate $G_\phi$ (and $\gef$). Our next concern is
how large $\gef$ can be. According to the general formula in
single-inflaton models to be derived in Sec.~\ref{sec:5} (see
(\ref{eq:g-eff-single})), it is at most $\sim m_{3/2}/m_\phi$.  In
fact, this is also true in the inflation models with multiple fields.
Therefore the decay rate (\ref{eq:decay-rate}) can be comparable to
that obtained by the decay via Planck-suppressed dimension $5$
operators. In other words, this direct gravitino production becomes
important especially when the total decay rate of the inflaton is
suppressed, i.e., the reheating temperature is low.  Therefore the
direct gravitino production via the interaction
(\ref{eq:inf2gravitino}) is complementary to the thermal gravitino
production which becomes more effective for higher $T_R$.
Fig.~\ref{fig:y32-tr} schematically shows this feature.  This specific
character enables us to put severe constraints on inflation models.

\subsection{Cosmological constraints on $\gef$}
In the following we assume that the reheating temperature satisfies
the bounds from the thermally produced gravitinos discussed in
Sec.~\ref{sec:3}~\footnote{
As pointed out in Ref.~\cite{Endo:2006tf}, the mixing between the
inflaton and the SUSY breaking field may enhance the reheating
temperature.  Including the effect of the mixing, we take the
reheating temperature as a free parameter throughout this paper.
}. The reheating temperature $T_R$ is related to the decay rate of the
inflaton into the SM particles (and their superpartners) by~\footnote{
We have assumed $\Gamma_{3/2} \ll \Gamma_{\rm SM}$, since the standard
cosmology would be upset otherwise.
}
\beq
\Gamma_{\rm SM} \simeq \lrfp{\pi^2 g_*}{10}{\frac{1}{2}} \frac{T_R^2}{M_P},
\label{eq:Trh-from-gp}
\eeq
where $g_* $ counts the relativistic degrees of freedom and hereafter
we set $g_* = 228.75$. When the Hubble parameter becomes comparable to
$\Gamma_{\rm SM}$, the inflaton decays. It is easy to see that $H \sim
\Gamma_{\rm SM} \ll m_{3/2}$ is realized at the decay, if the
reheating temperature satisfies the bounds from the gravitinos
produced by thermal scattering (i.e., (\ref{eq:rtemp_0}) $-$
(\ref{eq:unstable-g3}), (\ref{eq:const-from-lsp}), and
(\ref{eq:const-from-winolsp}) $-$ (\ref{eq:stable-g2})). Therefore the
inflaton decay into the gravitinos is effective.  The
gravitino-to-entropy ratio is then given by~\footnote{
Here we assume that the entropy comes solely from the
perturbative decay of the inflaton. 
}
\bea
Y_{3/2} &\simeq& 2 \,\frac{\Gamma_{3/2}}{\Gamma_{\rm SM}}\frac{3}{4}
		   \frac{T_R}{m_\phi},\non\\
             &\simeq& 4.5 \times 10^{5} \,|\gef|^2 
           \lrfp{m_{3/2}}{1{\rm\,TeV}}{-2}  	\lrfp{m_\phi}{10^{10}{\rm\,GeV}}{4} 
		\lrfp{T_R}{10^6{\rm\,GeV}}{-1},
\label{eq:ngs}
\eea
where we have neglected the gravitino production from the thermal
scattering.

First let us consider the cosmological bound on the gravitino
abundance for stable gravitinos of $m_{3/2} \lesssim 10{\rm GeV}$.
The gravitino abundance should not exceed the dark matter abundance;
\beq
m_{3/2} \,Y_{3/2} \;\leq\; \Omega_{\rm DM} \frac{\rho_c}{s} \;\lesssim\;
4.7 \times 10^{-10} {\rm\,GeV},
\label{eq:gra-abu}
\eeq
where $\rho_c$ is the critical density, and we used $\Omega_{\rm DM}
h^2 \lesssim 0.13$ at $95\%$ C.L. in the second inequality.  Combining
(\ref{eq:ngs}) and (\ref{eq:gra-abu}), we obtain
\beq
|\gef| \;\lesssim\; 3.2 \times 10^{-11}
           \lrfp{m_{3/2}}{1{\rm\,GeV}}{\frac{1}{2}} 	\lrfp{m_\phi}{10^{10}{\rm\,GeV}}{-2} 
		\lrfp{T_R}{10^6{\rm\,GeV}}{\frac{1}{2}},
\label{eq:const-on-G-stable}		
\eeq
for $T_R$ satisfying (\ref{eq:stable-g}) or (\ref{eq:stable-g2}).  To
further reduce this bound, we need to substitute the largest allowed
value of $T_R$ given by (\ref{eq:stable-g}) and
(\ref{eq:stable-g2}). Then we arrive at
\bea
\label{eq:const-on-Gstable0}
|\gef| &\lesssim &O(10^{-16})
           \lrfp{m_{3/2}}{1{\rm\,keV}}{\frac{1}{2}} 	\lrfp{m_\phi}{10^{10}{\rm\,GeV}}{-2}
\eea
for $m_{3/2} \simeq 1 {\rm\,keV} - 100{\rm\,keV}$, and
\bea	
|\gef| &\lesssim& 1.9 \times 10^{-10}  \lrfp{m_{\tilde{g}_3}}{500{\rm\,GeV}}{-1}
           \lrf{m_{3/2}}{1{\rm\,GeV}}	\,\lrfp{m_\phi}{10^{10}{\rm\,GeV}}{-2}
\label{eq:const-on-Gstable}
\eea
for $m_{3/2} \simeq 100 {\rm\,keV} - 10{\rm\,GeV}$.  It should be
noted that the constraints on $\gef$ become severer for lower $T_R$,
as clearly seen from (\ref{eq:const-on-G-stable}) (or Fig.~\ref{fig:y32-tr}).

Next we consider unstable gravitinos. The gravitino abundance is
severely constrained by BBN as discussed in Sec.~\ref{sec:3}.  We can
similarly derive the constraints on $\gef$ from (\ref{eq:rtemp_0}) $-$
(\ref{eq:unstable-Y2}), and (\ref{eq:ngs}):
\bea
   |\gef|  & \lesssim &  \left\{\begin{array}{lcl}
\ds{   (2-10)\times 10^{-12} 
   \lrfp{m_\phi}{10^{10}{\rm\,GeV}}{-2}}
    &{\rm for}&  m_{3/2}  \;\simeq\; 0.1 - 0.2~{\rm TeV} \\[0.8em]
\ds{    1 \times  10^{-11} 
   \lrfp{m_\phi}{10^{10}{\rm\,GeV}}{-2}}
   \,\, &{\rm for}& ~ m_{3/2}  \;\simeq\; 0.2 - 2~{\rm TeV} \\[0.8em]
\ds{     (2\times 10^{-11} - 2  \times  10^{-8} )
   \lrfp{m_\phi}{10^{10}{\rm\,GeV}}{-2}}
   \,\, &{\rm for}& m_{3/2}  \;\simeq\; 2 - 10~{\rm TeV} \\[0.8em]
\ds{    (0.6-6) \times 10^{-6} 
   \lrfp{m_\phi}{10^{10}{\rm\,GeV}}{-2}}
   \,\, &{\rm for}&  m_{3/2}  \;\simeq\; 10 - 30~{\rm TeV} 
   \end{array}\right.
\eea
for  $ B_h \simeq 1$, and
\bea
   |\gef|   & \lesssim  & \left\{\begin{array}{lcl}
 \ds{   (2\times10^{-12} - 6 \times10^{-9})
   \lrfp{m_\phi}{10^{10}{\rm\,GeV}}{-2}}
    &{\rm for}&  m_{3/2}  \;\simeq\; 0.1 - 1~{\rm TeV} \\[0.8em]
\ds{    6 \times 10^{-9}
    \lrfp{m_\phi}{10^{10}{\rm\,GeV}}{-2} } 
     &{\rm for}&  m_{3/2}  \;\simeq\; 1 - 3~{\rm TeV} \\[0.8em]
 \ds{    (0.1 - 2)  \times10^{-7}
    \lrfp{m_\phi}{10^{10}{\rm\,GeV}}{-2}  }
    &{\rm for}&  m_{3/2}  \;\simeq\; 3 - 10~{\rm TeV} 
    \end{array}\right. 
\eea
for $B_h \simeq 10^{-3}$.  For $m_{3/2}$ larger than $30 (10)$ TeV,
the constraint comes from the LSP abundance produced by the gravitino
decay. Using (\ref{eq:Yx-new}), (\ref{eq:const-from-lsp}), and
(\ref{eq:ngs}), we obtain
\bea
|\gef| \;\lesssim\; 5 \times 10^{-5}\,   \lrfp{m_{\rm LSP}}{100{\rm\,GeV}}{-1}
		 \lrf{m_{3/2}}{100{\rm\,TeV}} 	\lrfp{m_\phi}{10^{10}{\rm\,GeV}}{-2},
\eea
for $m_{3/2} \simeq 30 (10) - 100$ TeV. In particular, this can be
rewritten as
\bea
\label{eq:const-g-amsb}
|\gef| \;\lesssim\; 2 \times10^{-5} \, 	\lrfp{m_\phi}{10^{10}{\rm\,GeV}}{-2},
\eea
for the anomaly-mediated SUSY breaking with the wino LSP, where
we have used (\ref{eq:winomass}).

\begin{figure}[t]
\begin{center}
\includegraphics[width=10cm]{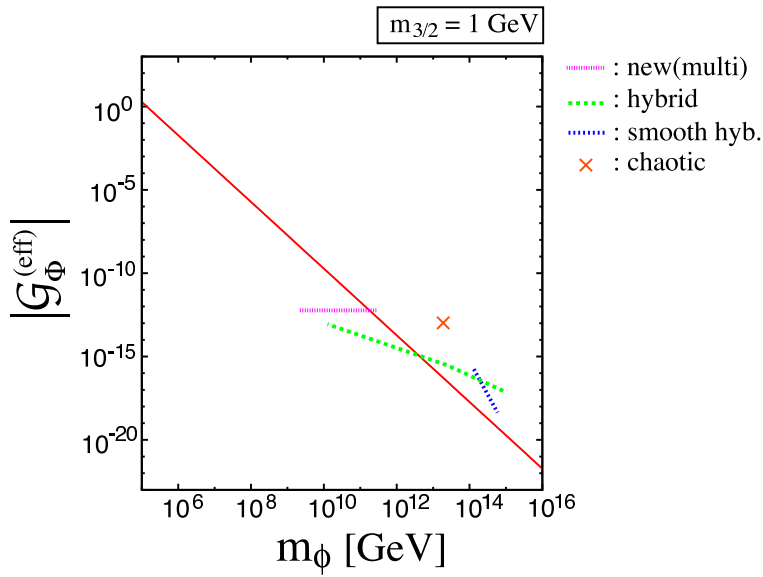}
\caption{Upper bound on the (effective) auxiliary field of the
inflaton $\gef$ as a function of the inflaton mass $m_\phi$, with
$m_{3/2} = 1{\rm\, GeV}$.  We set $m_{\tilde{g}_3} = 500$~GeV.  $T_R$
is set to be the largest allowed value, and the bound becomes severer
for lower $T_R$.  The typical values of $\gef$ and $m_\phi$ for the
multi-field new, hybrid, smooth hybrid, and chaotic inflation models
with $\kappa = 1$ are also shown.  The chaotic inflation can avoid
this bound by assuming $Z_2$ symmetry (see the text for details).  }
\label{fig:bound1GeV}
\end{center}
\end{figure}

\begin{figure}[t]
\begin{center}
\includegraphics[width=10cm]{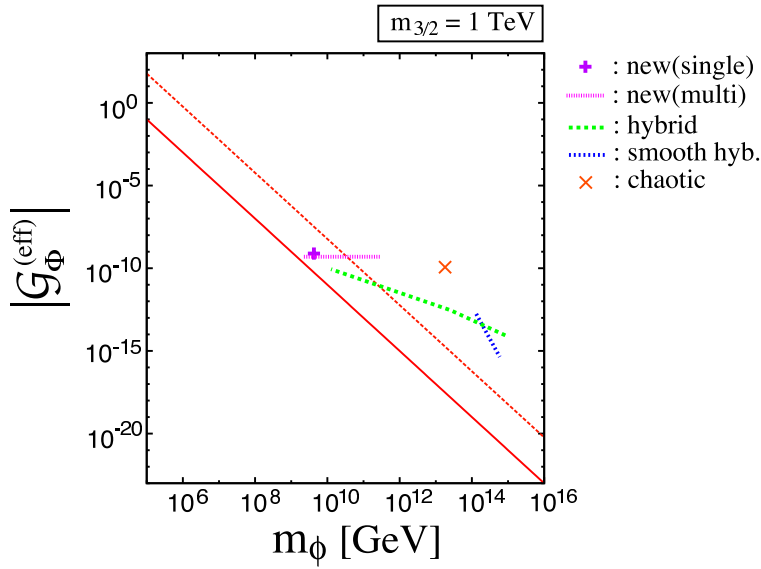}
\caption{ Same as Fig.~\ref{fig:bound1GeV} except for $m_{3/2} =
1{\rm\, TeV}$.  The typical values of $\gef$ and $m_\phi$ for the
single-field new inflation model with $\kappa = 1$ are also plotted.
The solid and dashed lines are for the hadronic branching ratio $B_h =
1$ and $10^{-3}$, respectively.  }
\label{fig:bound1TeV}
\end{center}
\end{figure}

\begin{figure}[t]
\begin{center}
\includegraphics[width=10cm]{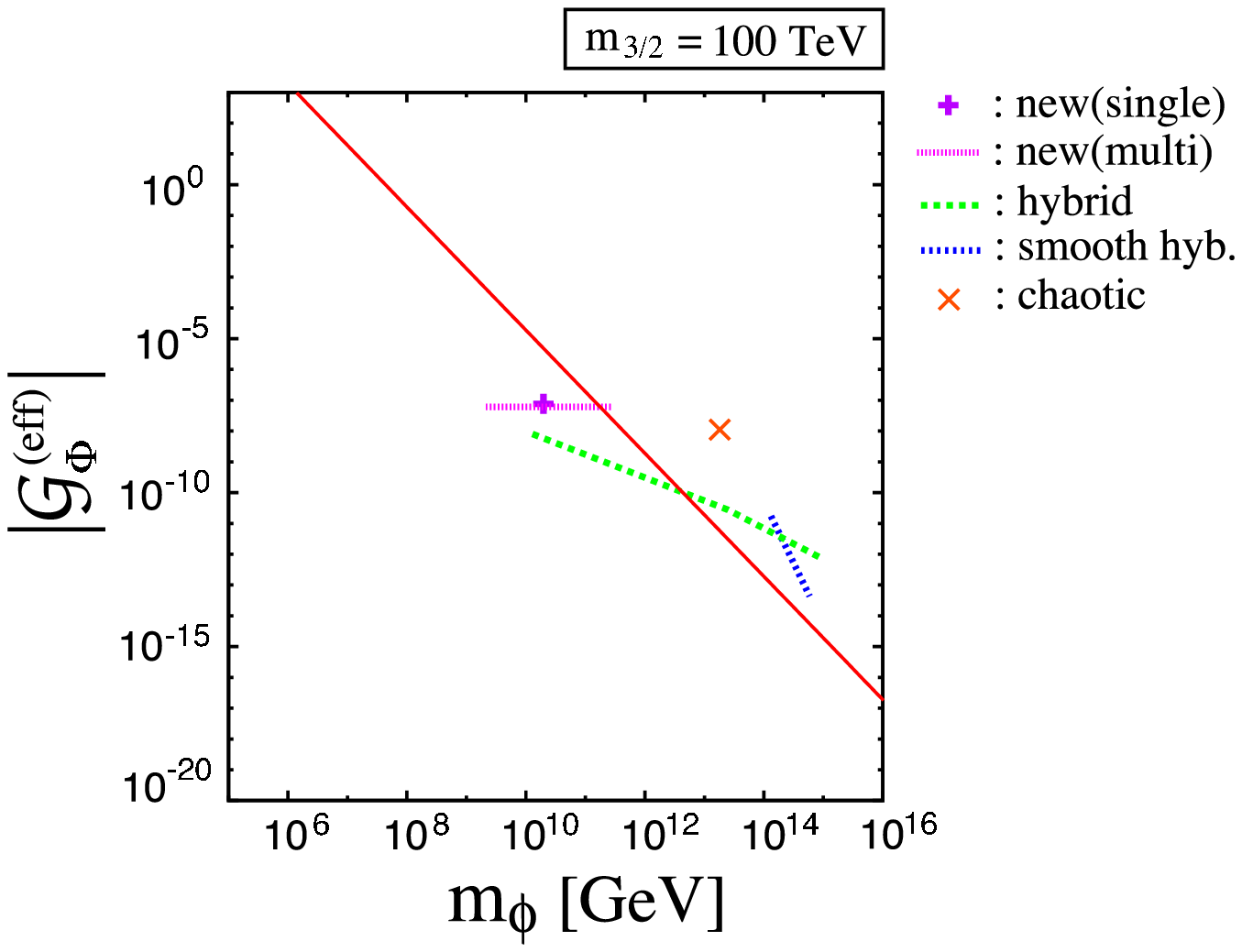}
\caption{Same as Fig.~\ref{fig:bound1TeV} except for $m_{3/2} = 100{\rm\, TeV}$.
}
\label{fig:bound100TeV}
\end{center}
\end{figure}

In Figs.~\ref{fig:bound1GeV} - \ref{fig:bound100TeV}, we show the
upper bounds on $\gef$ together with predictions of new, hybrid,
smooth hybrid, and chaotic inflation models to be derived in
Sec.~\ref{sec:5}, for representative values of the gravitino mass:
$m_{3/2} = 1{\rm\,GeV}$, $1$\,TeV, and $100$\,TeV, respectively.  From
these figures one can see that the bound is the severest in the case
of $m_{3/2} = 1$\,TeV due to the strict BBN bounds.  The bounds are
slightly relaxed for either (much) heavier or lighter gravitino mass.
Note that the constraints on inflation models do not change for the
stable gravitinos with $m_{3/2} \simeq 100{\rm\,keV} - 10{\rm\,GeV}$,
since both the upper bound and the actual value of $\gef$ in the
vacuum are proportional to $m_{3/2}$ (cf. (\ref{eq:const-on-Gstable})
and (\ref{eq:g-eff-single})).  The smooth hybrid inflation is excluded
for a broad region of the gravitino mass, unless $\kappa$ (see the
next section for the definition) is suppressed.  Similarly, for
$\kappa \sim O(1)$, a significant fraction of the parameter space in
the hybrid inflation model is excluded, and in particular, it is
almost excluded for $m_{3/2} = 1$\,TeV, while the new inflation is on
the verge of.  Even though the constraints on the hybrid inflation
model seems to be relaxed for smaller $m_\phi$, it is then somewhat
disfavored by WMAP three year data~\cite{Spergel:2006hy} since the
predicted spectral index approaches to unity. The chaotic inflation
model is also excluded unless $\kappa$ is suppressed due to some
symmetry.


\section{Explicit calculation of $\gef$ for several inflation models}
\label{sec:5}

In estimating the effective coupling of the inflaton $\phi$ with the
gravitino, the mixings with the SUSY breaking sector field is
important, as pointed in Ref.~\cite{Dine:2006ii} for specific
cases. In fact one can rigorously estimate the coupling in a rather
generic way~\cite{Endo:2006tf}.  In this section, we would like to
show how to obtain $\gef$, based on the argument of
Ref.~\cite{Endo:2006tf}.

The point is that the inflaton field $\phi$ does not coincide with the
mass eigenstate after inflation due to the mixings with the SUSY
breaking sector field $z$. There are three sources for the mixings:
(i) kinetic terms (or equivalently, the K\"ahler metric); (ii)
non-analytic (NA) mass terms; (iii) analytic (A) mass terms. Although
the mixing in the K\"ahler metric can be important for the reheating
processes~\cite{Endo:2006tf}, we neglect it here since it does not
affect the coupling with the gravitinos. In the following we focus on
the mixings in the mass terms.

\subsection{Single-field inflation model}
\label{sec:single}
Let us first consider a single-field inflation model, with the K\"aler
metric $g_{ij^*} = \delta_{ij}$. In the Einstein frame, the SUGRA
Lagrangian contains the scalar potential, $V = e^G (G^i G_i - 3)$. The
non-analytic (NA) and analytic (A) mass terms are written as
\begin{eqnarray}
  M_{ij*}^2 &=& \frac{\partial^2 V}{\partial \varphi^i \partial \varphi^{\dagger j}}
  \;=\; e^G \left( 
    \nabla_i G_k \nabla_{j*} G^k - R_{ij^*k\ell^*} G^k G^{\ell*} + g_{ij^*}
  \right),
  \label{eq:Mijs}
  \\
  M_{ij}^2 &=& M_{ji}^2 
  \;=\; \frac{\partial^2 V}{\partial \varphi^i \partial \varphi^j}
  \;=\; e^G \left( 
    \nabla_i G_j + \nabla_j G_i + G^k \nabla_i \nabla_j G_k 
  \right),
  \label{eq:Mij}
\end{eqnarray}
respectively, where we have assumed the vanishing cosmological
constant, $G^i G_i = 3$, and used the potential minimization
condition, $G^i \nabla_k G_i + G_k = 0$ in the vacuum. The gravitino
mass is given by $m_{3/2} = \la e^{G/2} \ra$.  Here $R_{ij^*k\ell^*}$
is the curvature of the K\"ahler manifold, defined by $R_{ij^*k\ell^*}
= g_{ij^*k\ell^*} - g^{mn^*} g_{mj^*\ell^*} g_{n^*ik}$.  Also the
covariant derivative of $G_i$ is defined by $\nabla_i G_j = G_{ij} -
\Gamma^k_{ij} G_k$, where the connection, $\Gamma_{ij}^k = g^{k\ell^*}
g_{ij\ell^*}$, and $\nabla_k g_{ij^*} = 0$ is satisfied.

We assume that the inflaton $\phi$ is heavy due to a large
supersymmetric mass, $m_\phi \equiv |e^{G/2} \nabla_\phi G_\phi| \gg
m_{3/2}$, and that $M^2_{\phi \bar \phi}$ dominates over the other elements
of the mass terms. Then the NA mass terms can be diagonalized by the
following transformation:
\begin{eqnarray}
  \Phi &\equiv& \phi + \epsilon z,\non\\
   Z &\equiv& z - \epsilon^* \phi,
\end{eqnarray}
where $\epsilon$ represents the mixing angle. Here we have assumed
$|\epsilon| \ll 1$ and neglected those terms of $O(\epsilon^2)$.
Since $M_{\phi \bar\phi}^2$ dominates over the other components in the
mass matrix, the mixing angle is given by the ratio of $M_{\phi
\bar\phi}^2$ to the off-diagonal component:
\beq
\label{eq:epsilon}
\epsilon \simeq  \frac{M_{z \bar \phi}^2}{M_{\phi \bar \phi}^2}.
\eeq
As emphasized in Ref.~\cite{Endo:2006tf}, the NA mass eigenstates
$(\Phi, Z)$ do not necessarily coincide with the true mass
eigenstates. In fact, the analytic mass terms generically provide
further mixing between $\phi$ and $z^\dag$.  The true mass eigenstates
are therefore
\begin{eqnarray}
\label{eq:a-phi}
  \tilde \Phi &\equiv& \phi+ \epsilon z + \tilde \epsilon z^\dagger, \\
\label{eq:a-z}
  \tilde Z    &\equiv& z- \epsilon^* \phi - \tilde \epsilon \phi^\dagger,
\end{eqnarray}
where the mixing angle $\tilde \epsilon$, which is assumed to be much
smaller than unity, is given by
\beq
\label{eq:tilep}
\tilde \epsilon \;\simeq\; \frac{M_{\bar \phi \bar z}^2}{M_{\phi \bar\phi}^2}.
\eeq
Below we show that the coupling with the gravitinos is suppressed in
the NA mass eigenstates, but it is not the case in the true mass
eigenstates if the K\"ahler potential is non-minimal.

In the NA mass eigenstates, the off-diagonal element of the
non-analytic mass term is zero by definition:
\beq
M^2_{\Phi \bar Z} = e^G \left(\nabla_\Phi G_\Phi \nabla_{\bar Z} G_{\bar \Phi }
+ \nabla_\Phi G_Z \nabla_{\bar Z} G_{\bar Z} - R_{\Phi  
\bar Z i j^*}G^i G^{j^*} \right) =0,
\eeq
which leads to
\beq
\label{eq:Gzphi}
\nabla_{\bar Z} G_{\bar \Phi } \;\simeq\; \frac{R_{\Phi \bar Z i  
j^*}G^i G^{j^*} }{ \nabla_\Phi G_\Phi },
\eeq
where we have used $|\nabla_\Phi G_\Phi| \gg |\nabla_Z G_Z|$.  On the
other hand, the potential minimization condition for $\Phi$ reads
\beq
G_{\bar \Phi} \nabla_\Phi G_\Phi + G_{\bar Z} \nabla_\Phi G_Z  
+ G_\Phi = 0,
\eeq
which can be solved for $G_\Phi$:
\beq
\label{eq:Gphi}
G_\Phi  \;\simeq\; -  \frac{\nabla_{\bar \Phi} G_{\bar Z}}{ \nabla_ 
{\bar \Phi} G_{\bar \Phi}} G_{Z}.
\eeq
Substituting (\ref{eq:Gzphi}) into (\ref{eq:Gphi}),
we arrive at
\beq
\label{eq:gphi-na}
|G_\Phi| \;\simeq\; 3\sqrt{3}  \frac{|R_{\Phi \bar Z Z \bar Z}| }{|  
\nabla_{\bar \Phi} G_{\bar \Phi}|^2},
\eeq
where we have used $|G_Z| = |G^Z| \simeq \sqrt{3}$.  Thus $G_\Phi$ is
always proportional to $m_{3/2}^2/m_\phi^2 \ll 1$.  For the minimal
K\"ahler potential, $G_\Phi$ is exactly zero in this basis.  In the
mass-eigenstate basis, therefore, the effective coupling of the
inflaton $\tilde \Phi$ with the gravitinos dominantly comes from the
mixing in the analytic mass terms:
\beq
\label{eq:g-eff-single}
|\gef| \;\simeq\;  \left|\tilde \epsilon \,G_z \right| \simeq 3 |g_{\bar \phi z z}| \frac{m_{3/2}}{m_{\phi}}.
\eeq
For instance, let us consider $\delta K = (1/2)\kappa|\phi|^2 zz +
{\rm h.c.}$, which is expected to be present if $z$ is singlet under
any symmetries.  For the non-minimal K\"ahler potential, the
effective coupling becomes
\beq
\label{eq:kappa}
|\gef| \;\simeq\; 3 \kappa \la \phi \ra \frac{m_{3/2}}{m_\phi},
\eeq
where $\la \phi \ra$ denotes the vacuum expectation value (VEV) of
$\phi$.  Therefore, $\gef$ is proportional to $G_\phi \sim \la \phi \ra
m_{3/2}/m_\phi$~\cite{KTY-I}, for this choice of the interaction
between $\phi$ and $z$.

Here let us comment on the case that the mass of the SUSY breaking
field $\tilde Z$, $m_z$, is larger than $m_\phi$ due to the non-SUSY
mass term. Such situation may be realized in the dynamical SUSY
breaking models~\cite{Witten:1981nf}. Then the $|\gef|$ becomes of
$O(\la \phi \ra m_{3/2}/m_\phi)$ even if the K\"ahler potential is
minimal~\cite{Dine:2006ii,Endo:2006tf}.  To be conservative, however,
we assume that $m_\phi > m_z$ in the following discussion.

As a concrete example, here we study the new inflation
model~\cite{Kumekawa:1994gx,Izawa:1996dv,Ibe:2006fs}.  In the new
inflation model, the K\"ahler potential and superpotential of the
inflaton sector are written as
\bea
K(\phi,\phi^\dag) &=& |\phi|^2 + \frac{k}{4} |\phi|^4,\non\\
W(\phi) &=&  v^2\phi -  \frac{g}{n+1}\, \phi^{n+1}.
\eea
where the observed density fluctuations are explained for $v = 4\times
10^{-7} \, (0.1/g)^{1/2}$ and $k \lesssim 0.03$ in the case of
$n=4$~\cite{Ibe:2006fs}.  After inflation, the inflaton $\phi$ takes
the expectation value $\la\phi\ra\simeq (v^2/g)^{1/n}$. In this model
the inflaton mass is given by $m_{\phi} \simeq n v^2/\la\phi\ra$, and
the gravitino mass is related to $v$ as $m_{3/2} \simeq n v^2
\la\phi\ra /(n+1)$, since the inflaton induces the spontaneous
breaking of the $R$-symmetry. Thus, (\ref{eq:g-eff-single}) leads
to~\footnote{
The relation (\ref{eq:g-eff-single}) remains virtually unchanged in
the presence of the quartic coupling in the K\"ahler potential.
}
\beq
|\gef |\simeq \frac{3}{n+1} |g_{\bar \phi z z}| \lrfp{v^2}{g}{\frac{2}{n}}.
\eeq
For the interaction $\delta K = (1/2)\kappa|\phi|^2 zz + {\rm h.c.}$,
this becomes
\beq
|\gef |\simeq \frac{3 \kappa }{n+1} \lrfp{v^2}{g}{\frac{3}{n}}.
\eeq
In the case of $n=4$, $|\gef| \simeq 8 \times 10^{-10} \kappa$ and
$m_\phi \simeq 4 \times 10^9$ GeV for $m_{3/2}=1$~TeV, while $|\gef|
\simeq 8\times 10^{-8}\kappa$ and $m_\phi \simeq 2 \times 10^{10}$ GeV
for $m_{3/2}=100$~TeV.  Note that $m_{3/2} \ll 1\,$TeV cannot be
realized unless $g \gg 1$.  We plot these results with $\kappa = 1$ in
Figs.~\ref{fig:bound1TeV} and \ref{fig:bound100TeV}. We can see that
the new inflation model is on the verge of being excluded for $m_{3/2}
= 1$ TeV~\footnote{It may survives if $\kappa \simeq 10^{-2}$ as
suggested in the large-cutoff SUGRA~\cite{Ibe:2004mp}. We thank M. Ibe
and Y. Shinbara for useful discussion.}, while it is close to but
slightly below the bound for $m_{3/2} = 100$~TeV.  This single-field
new inflation model will be discussed in detail in Ref.~\cite{ISY}.

\subsection{Multiple-field inflation model}

Next we consider an inflation model with multiple fields, for which
the formula (\ref{eq:g-eff-single}) cannot be simply applied as it
is. Although we generically need to evaluate $\gef$ for each inflation
model, there is an important class of models described by the
following superpotential:
\beq
\label{eq:multi-generic}
W(\phi,\psi)\;=\; \phi f(\psi),
\eeq
where $f(\psi)$ is a function of $\psi$.  The potential minimum in the
global SUSY limit is located at
\bea
\la \phi \ra&=&0,\non\\
\la \psi \ra &=& \psi_0,
\label{eq:min-global-generic}
\eea
where $\psi_0$ satisfies $f(\psi_0)=0$.  Note that the true minimum is
slightly displaced from (\ref{eq:min-global-generic}), once the SUSY
breaking field is taken into account~\cite{KTY-I,Dvali:1997uq}.

For instance, the above class of the models includes a new inflation
model~\cite{Asaka:1999jb} and a hybrid inflation
model~\cite{Copeland:1994vg,Dvali:1994ms,Linde:1997sj}, described by
\beq
\label{eq:multi}
W(\phi,\psi)\;=\; \phi \left(\mu^2 - \frac{\psi^n}{M^{n-2}} \right),
\eeq
where $\mu$ determines the inflation energy scale and $M$ is an
effective cut-off scale. In the new inflation model $\psi$ plays a
role of the inflaton, while $\phi$ is the inflaton in the hybrid
inflation model.

The inflaton fields $\phi$ and $\psi$ have almost the same masses,
\beq
\label{eq:susy-mass-2field}
m_\phi \;\simeq\; m_\psi \;\simeq\; \left| e^{G/2} \nabla_\phi G_\psi \right|,  
\eeq
which are assumed to be much larger than the gravitino mass.  It
should be noted that $\phi$ and $\psi$ (and/or $\psi^\dag$) almost
maximally mix with each other to form the mass eigenstates due to the
almost degenerate masses.  To see this let us take the NA
mass-eigenstate basis $(\Phi, \Psi, Z)$ in which the non-analytic mass
matrix is diagonalized except for $\phi-\psi$ mixing. The difference
between the diagonal components of the non-analytic mass matrix is
small: $|M^2_{\Phi \bar \Phi}-M^2_{\Psi \bar \Psi}| = O(m_{3/2}^2)$,
while the off-diagonal component in the analytic mass matrix is
relatively large: $M^2_{\Phi \Psi} = O(m_{3/2} m_\phi )$~\footnote{ In
addition, the off-diagonal component in the non-analytic mass matrix
as well can be as large as $M^2_{\Phi \bar \Psi} = O(m_{3/2} m_\phi )$
if $f''(\psi_0) \sim f'(\psi_0)/\psi_0$, and the mixing is almost
maximal in this case too.  }, resulting in the almost maximal mixing
between $\phi$ and $\psi^\dag$.  This mixing is effective at the
inflaton decay, since the Hubble parameter at the decay should be
(much) smaller than $O(m_{3/2})$ to satisfy the bounds from
the thermally produced gravitinos.  However, since the mixing is due
to the specific character of (\ref{eq:multi-generic}) and it occurs
within the inflaton sector, we leave it for a moment. Then we can
similarly show that the auxiliary fields $G_\Phi$ and $G_\Psi$ are
proportional to $m_{3/2}^2/m_\phi^2$ in the NA mass eigenstates
$(\Phi, \Psi, Z)$. Therefore the effective couplings with the
gravitinos arise mainly from the mixings in the analytic mass terms,
as in the single-field inflation~\footnote{
Note that the dependence of the right-handed side on $\phi$ and $\psi$
originates from the SUSY mass (\ref{eq:susy-mass-2field}), which is
peculiar to the form of the superpotential (\ref{eq:multi}).
}:
\bea
|\gef| &\simeq& 3 |g_{\bar \psi z z}| \frac{m_{3/2}}{m_\phi},\non\\
|\gep|&\simeq& 3 |g_{\bar \phi z z}| \frac{m_{3/2}}{m_\psi}.
\label{eq:eff-g-multi}
\eea
For such interactions as $\delta K = (\kappa/2) |\psi|^2 z z + (\tilde
\kappa/2) |\phi|^2 z z +{\rm h.c.}$, we have
\bea
|\gef| &\simeq& 3 \kappa \la \psi \ra \frac{m_{3/2}}{m_\phi},\non\\
|\gep|&\simeq& 3 \tilde \kappa \la \phi \ra\frac{m_{3/2}}{m_\psi}.
\eea
Therefore $|\gep|$ is suppressed compared to $|\gef|$ if $\la \psi \ra
\gg \la \phi \ra$ as in the case of  (\ref{eq:multi}).

The true mass eigenstates are obtained after taking account of the
(almost) maximal mixing between $\phi$ and $\psi (\psi^\dag)$ discussed above:
\beq
\varphi_{\pm} \;\simeq \frac{\phi \pm \psi^{(\dag)}}{\sqrt{2}},
\eeq
where we have omitted the relatively small mixings with $z$ for
simplicity, but they are included in the definition of
$\varphi_\pm$. For $|\gef| \gg |\gep|$, the effective couplings of
$\varphi_\pm$ with the gravitinos are roughly given by
\beq
\label{eq:max-mix}
|{\cal G}^{\rm (eff)}_{\varphi_\pm}| \;\simeq\; \frac{1}{\sqrt{2}} |\gef|.
\eeq

\subsubsection{New inflation model}
The new inflation discussed in Sec.~\ref{sec:single} is also realized
for~\cite{Asaka:1999jb}
\bea
K  & = & |\phi|^2 + |\psi|^2 + \frac{k_1}{4}|\phi|^4
+ k_2 |\phi|^2 |\psi|^2 + \frac{k_3}{4}|\psi|^4,\non\\
W & = & \phi (v^2 -g \,\psi^4),
\eea
in which the inflaton is $\psi$, while $\phi$ stays at the origin
during and after inflation~\footnote{If one introduces a constant term
in the superpotential, the $\phi$ shifts from the origin.}. If one
defines $k \equiv k_2 -1$, the scalar potential for the inflaton
$\psi$ becomes the same as the single-field new inflation model,
although the gravitino mass is not related to the inflaton parameters.
After the inflation ends, the energy of the universe is dominated by
the oscillation energy of $\psi$~\footnote{
The tachyonic preheating~\cite{Felder:2000hj,Desroche:2005yt} is known
to occur in this model, and if it occurs, the homogeneous mode of the
inflaton $\psi$ disappears soon and the excited $\psi$ particles are
produced. This instability itself does not relax the
gravitino-overproduction problem, since these $\psi$ particles will
decay perturbatively into the SM particles and their superpartners.
Further, if the $\psi$ particles are relativistic, the decay is
delayed, making the problem even worse.
}. Although $\gep$ is suppressed compared to $\gef$, the effective
coupling to the gravitinos is given by (\ref{eq:max-mix}), since
$\phi$ and $\psi$ almost maximally mixes with each other in the
vacuum.  Thus the constraint on this model is comparable to that on
the single-field new inflation. For the non-minimal coupling $\delta K
= (1/2)\kappa|\psi|^2 zz + {\rm h.c.}$, the effective coupling to the
gravitinos is given by
\bea
|{\cal G}^{\rm (eff)}_{\varphi_\pm}|  &\simeq& \frac{3}{\sqrt{2}}
 \kappa \la \psi \ra \frac{m_{3/2}}{m_\phi}.
\eea
We plot the value of $|{\cal G}^{\rm (eff)}_{\varphi_\pm}|$ for $g =
10^{-4} - 1$ and $k=10^{-4}-10^{-1.5}$ with the e-folding number
$N=50$ in Figs.~\ref{fig:bound1GeV}, \ref{fig:bound1TeV}, and
\ref{fig:bound100TeV}.  Thus the (multi-field) new inflation model
is on the verge of being excluded, if $\kappa$ is order unity.

\subsubsection{Hybrid inflation model}
The hybrid inflation model contains two kinds of superfields: one is
$\phi$ which plays a role of inflaton and the others are waterfall
fields $\psi$ and
$\tilde{\psi}$~\cite{Copeland:1994vg,Dvali:1994ms,Linde:1997sj}.
After inflation ends, $\phi$ as well as $\psi$($\tilde{\psi}$)
oscillates around the potential minimum and dominates the universe
until the reheating.

The superpotential $W(\phi, \psi,\tilde{\psi})$ for the inflaton
sector is
\begin{equation}
   \label{eq:spot_hyb}
   W(\phi, \psi,\tilde{\psi}) = \phi (\mu^{2}  
   - \lambda \tilde{\psi}\psi),
\end{equation}
where $\psi$ and $\tilde \psi$ are assumed to be charged under $U(1)$
gauge symmetry.  Here $\lambda$ is a coupling constant and $\mu$ is
the inflation energy scale. The potential minimum is located at
$\la\phi\ra = 0$ and $\langle \psi\rangle = \langle\tilde{\psi}\rangle
= \mu/\sqrt{\lambda}$ in the SUSY limit.  For a successful inflation,
$\mu$ and $\lambda$ are related as $\mu \simeq 2\times
10^{-3}\lambda^{1/2}$ for $\lambda \gtrsim 10^{-3}$, and $\mu \simeq
2\times 10^{-2}\lambda^{5/6}$ for $\lambda \lesssim 10^{-3}$.
Moreover, in this type of hybrid inflation there exists a problem of
cosmic string formation because $\psi$ and $\tilde\psi$ have $U(1)$
gauge charges. To avoid the problem the coupling $\lambda$ should be
small as, $\lambda \sim 10^{-4}$~\cite{Endo:2003fr}.

Due to the D-term potential one linear combination of $\psi$ and
$\tilde \psi$, given by $\psi^{(-)} \equiv
(\psi-\tilde{\psi})/\sqrt{2}$, has a large mass of $\sim g \la \psi
\ra$ ($g$ denotes the gauge coupling), while the other, $\psi^{(+)}
\equiv (\psi+\tilde{\psi})/\sqrt{2}$ has a mass equal to that of
$\phi$: $m_{\psi^{(+)}}=m_\phi = \sqrt{2} \lambda \langle \psi
\rangle$.  It is the latter that (almost) maximally mixes with $\phi$
to form mass eigenstates.  Since the form of the superpotential is
almost identical to (\ref{eq:multi}), it is straightforward to extend
the results (\ref{eq:eff-g-multi}) and (\ref{eq:max-mix}) to obtain
\begin{eqnarray}
\label{eq:G_phi_hyb}
 |{\cal G}^{\rm (eff)}_{\varphi_\pm}| &\simeq& \frac{3}{\sqrt{2}} \kappa (\sqrt{2}\langle \psi \rangle)
 \frac{m_{3/2}}{m_{\phi}},
 \end{eqnarray}
for the non-minimal coupling $\delta K = (1/2)\kappa|\psi^{(+)}|^2 zz
+ {\rm h.c.}$.  Note that VEV of $\psi^{(+)}$ is equal to
$\sqrt{2}\langle \psi \rangle$.

For $\lambda \sim 10^{-1} -10^{-5}$~\cite{Bastero-Gil:2006cm} we
obtain $\mu \sim 8\times 10^{-4} - 1\times10^{-6}$, $ |{\cal G}^{\rm
(eff)}_{\varphi_\pm}| \sim 9\times 10^{-15} \kappa
(m_{3/2}/1{\rm\,TeV})- 9\times 10^{-11} \kappa (m_{3/2}/1{\rm\,TeV})$
and $m_{\phi} \sim 8\times 10^{14} - 1 \times 10^{10} $~GeV.  From
Fig.~\ref{fig:bound1TeV}, one can see the hybrid inflation model is
almost excluded by the gravitino overproduction for $m_{3/2} =
1{\rm\,TeV}$, if $\kappa$ is order unity.  For $m_{3/2} = 1{\rm\,GeV}$
and $100\,$TeV, the constraints become slightly mild, but a
significant fraction of the parameter space is still excluded (see
Figs.~\ref{fig:bound1GeV} and \ref{fig:bound100TeV}).  Although the
constraints on $|{\cal G}^{\rm (eff)}_{\varphi_\pm}|$ become relaxed
for smaller $m_\phi$ (i.e., smaller $\lambda \lesssim 10^{-4}$), it is
then somewhat disfavored by the WMAP data~\cite{Spergel:2006hy} since
the density fluctuation becomes almost scale-invariant.

Next let us consider a smooth hybrid inflation
model~\cite{Lazarides:1995vr}, which predicts the scalar spectral index as
$n_s \simeq 0.97$, which is slightly smaller than
the simple hybrid inflation model.  The superpotential of the inflaton
sector is
\beq
 \label{eq:spot_smhyb}
   W(\phi, \psi,\tilde{\psi}) = \phi \left(\mu^{2}  
   - \frac{ (\tilde{\psi}\psi)^n}{M^{2n-2}}\right).
\eeq
The VEVs of $\psi$ and $\tilde \psi$ are given by $\la \psi \ra = \la
\tilde \psi \ra=(\mu M^{n-1})^{1/n}$, and we assume that $\psi =
\tilde{\psi}$ always holds due to the additional D-term potential.
Then one of the combination, $\psi^{(+)}\equiv
(\psi+\tilde{\psi})/\sqrt{2}$, mixes with $\phi$, and the effective
coupling with the gravitinos is given by
\bea
\label{eq:G_phi_smhyb}
 |{\cal G}^{\rm (eff)}_{\varphi_\pm}| &\simeq& \frac{3}{\sqrt{2}} \kappa (\sqrt{2}\langle \psi \rangle)
 \frac{m_{3/2}}{m_{\phi}},
\eea
for $\delta K = (1/2)\kappa|\psi^{(+)}|^2 zz + {\rm h.c.}$.  Here we
have defined $m_\phi = \sqrt{2}n \mu^2/\la \psi \ra$.  The constraint
on the model is more or less similar to that on the hybrid inflation
model.  In fact, for $n=2$ we obtain $\mu \sim 4\times 10^{-4} -
9\times10^{-5}$, $ |{\cal G}^{\rm (eff)}_{\varphi_\pm}| \sim 2\times
10^{-13}\kappa(m_{3/2}/1{\rm\,TeV})- 4\times 10^{-16}\kappa
(m_{3/2}/1{\rm\,TeV})$ and $m_{\phi} \sim 1\times 10^{14} - 6 \times
10^{14} $~GeV.  From Figs.~\ref{fig:bound1GeV} - \ref{fig:bound100TeV}
one can see that the smooth hybrid inflation model is excluded for a
broad range of $m_{3/2}$ for $\kappa = O(1)$.

Lastly let us comment on the D-term inflation model~\cite{Binetruy:1996xj}, in 
which one of the waterfall fields, $\psi_{-}$, obtains a large VEV of $\sim 10^{16}{\rm\,GeV}$. 
The $\psi_{-}$ field can decay into a pair of the gravitino if there is a coupling like 
$|\psi_{-}|^2 zz / 2 +{\rm h.c.}$ in the K\"ahler potential. However, after inflation, the 
universe is dominated by the two fields: one is the $\psi_{-}$ field and the other is 
the inflaton, $S$. The resultant gravitino abundance thus depends on both the reheating processes
of these two fields and the relative portion of the energy in each field~\cite{Kolda:1998kc}. 
Therefore we cannot put a rigorous bound on the D-term inflation model.

\subsubsection{Chaotic inflation model}
A chaotic inflation \cite{Linde:1983gd} is realized in SUGRA, based on
a Nambu-Goldstone-like shift symmetry of the inflaton chiral multiplet
$\phi$ ~\cite{Kawasaki:2000yn,Kawasaki:2000ws}. Namely, we assume that
the K\"ahler potential $K(\phi,\phi^\dag)$ is invariant under the
shift of $\phi$,
\begin{equation}
  \phi \rightarrow \phi + i\,A,
  \label{eq:shift}
\end{equation}
where $A$ is a dimensionless real parameter. Thus, the K\"ahler
potential is a function of $\phi + \phi^\dag$; $K(\phi,\phi^\dag) =
K(\phi+\phi^\dag)= c\,(\phi+\phi^\dag) + \frac{1}{2}
(\phi+\phi^\dag)^2 + \cdots$, where $c$ is a real constant and must be
smaller than $O(1)$ for a successful inflation.  We will identify its
imaginary part with the inflaton field $\varphi \equiv \sqrt{2} {\rm
\,Im}[\phi]$.  Moreover, we introduce a small breaking term of the
shift symmetry in the superpotential in order for the inflaton
$\varphi$ to have a potential:
\begin{equation}
  W(\phi,\psi) = m\,\phi \,\psi, 
  \label{eq:mass}
\end{equation}
where we introduced a new chiral multiplet $\psi$, and $m \simeq
10^{13}$GeV determines the inflaton mass.

The scalar potential is given by
\bea 
V(\eta, \varphi, \psi) &=& m^2 e^{K} \left[|\psi|^2 \left(1+2
\left(\eta + \frac{c}{\sqrt{2}}\right)\eta + \left(\eta +
\frac{c}{\sqrt{2}}\right)^2(\eta^2+\varphi^2)\right)\right.\non\\ &&
\left.~~~~~~~~~+\frac{1}{2}(\eta^2+\varphi^2)(1-|\psi|^2+|\psi|^4)\right]
\eea
with \beq K \;=\; \left(\eta +
\frac{c}{\sqrt{2}}\right)^2-\frac{c^2}{2}+|\psi|^2, \eeq where we have
assumed the minimal K\"ahler potential for $\psi$, and defined $\eta
\equiv \sqrt{2} {\rm\,Re}[\phi]$.  Note that $\eta$ and $\psi$ cannot be
larger than the Planck scale, due to the prefactor $e^K$.  On the
other hand, $\varphi$ can be larger than the Planck scale
~\cite{Kawasaki:2000yn}, since $\varphi$ does not appear in $K$.  For
$\varphi \gg 1$, $\eta$ acquires the mass comparable to the Hubble
parameter and quickly settles down to the minimum, $\eta \simeq
-c/\sqrt{2}$. Then the scalar potential during inflation is given by
\beq
V(\eta,\varphi,\psi) \;\simeq\; \frac{1}{2}m^2 \varphi^2 + m^2 |\psi|^2.
\eeq
For $\varphi \gg 1$ and $|\psi| < 1$, the $\varphi$ field dominates the
potential and the chaotic inflation takes place (for details see
Refs~\cite{Kawasaki:2000yn,Kawasaki:2000ws}).

The effective auxiliary field of $\psi$ is given by
\beq
|{\cal G}^{\rm (eff)}_{\Psi}| \simeq 3 g_{\bar \phi z z} \frac{m_{3/2}}{m} =
3 \kappa \frac{m_{3/2}}{m},
\eeq
where we have assumed the non-minimal coupling $\delta K = (1/2)\kappa
(\phi+\phi^\dag) zz + {\rm h.c.}$ in the second equality.  This K\"ahler
potential is invariant under the shift symmetry (\ref{eq:shift}).
Note that $|{\cal G}^{\rm (eff)}_{\Phi}|$ is suppressed for e.g.,
$\delta K = (1/2)\tilde \kappa |\psi|^2 zz + {\rm h.c.}$ due to $\la
\psi \ra \ll 1$.  Taking account of the mixing between $\phi$ and
$\psi^\dag$, the effective coupling with the gravitinos is given by
\beq
|{\cal G}^{\rm (eff)}_{\varphi_\pm}| \simeq 
\frac{3}{\sqrt{2}} \kappa \frac{m_{3/2}}{m}.
\label{eq:chaotic}
\eeq
It is worth noting that both real and imaginary components of $\phi$
can decay into a pair of the gravitinos via the mixings with $z$ and
$\psi$. One might suspect that it is only the real component of $\phi$
that can decay into the gravitinos, since the shift symmetry dictates
that the only real component $(\phi + \phi^\dag)$ appears in the
K\"ahler potential.  However, it is not surprising that this is not
the case, since the enhanced decay amplitude is proportional to powers
of the large SUSY mass $m$ that explicitly violates the shift
symmetry.

We plot the result (\ref{eq:chaotic}) with $\kappa = 1$ in
Figs.~\ref{fig:bound1GeV}, \ref{fig:bound1TeV}, and
\ref{fig:bound100TeV}.  Although the coupling is too large if $\kappa
= O(1)$, it should be noted that in this chaotic inflation model we
can realize $|{\cal G}^{\rm (eff)}_{\varphi_\pm}| \simeq 0$ by
assuming an approximate $Z_2$ symmetry.  Therefore, the new gravitino
problem does not exist in this case.  A detailed discussion on the
chaotic inflation model will be given in \cite{kty-3}.

\section{Conclusions}
\label{sec:6}

Throughout this paper we have assumed no entropy production late after
the reheating of inflation. We briefly discuss potential problems when
a late-time entropy production~\cite{Lyth:1995ka} occurs.  First of
all, the cosmological constraints on the reheating temperature shown
in Sec.~\ref{sec:3} would be relaxed and the Hubble parameter at the
inflaton-decay time does not necessarily satisfy the condition
$H<m_{3/2}$ for the formula~(\ref{eq:ngs}) to be applicable.  Thus,
the cosmological constraints on $\gef$ would become milder
\footnote{Even if the reheating temperature is higher than the
cosmological bounds discussed in Sec.~\ref{sec:single}, the direct
gravitino production by inflaton decays can occur and the
formula~(\ref{eq:decay-rate}) is applicable as long as the condition
$H<m_{3/2}$ is satisfied at the decay time. In this case, we must
consider the direct production with a great caution, since it
dominates over the thermal production if
$|\gef|>10^{-13}(T_R/10^{10}{\rm GeV})(10^{13}{\rm GeV}/m_\phi)^2
(m_{3/2}/1{\rm TeV})$.}.  On the other hand, we must be careful about
the gravitino production in decay processes of the field $X$
responsible for the late-time reheating. One may have a similar
stringent constraint on ${\cal G}^{\rm (eff)}_X$. An obvious way to
induce late-time entropy production avoiding the problem is to assume
the late-time decay of a scalar field with a mass smaller than $2
m_{3/2}$.  In addition, there is another interesting example that is
free from the problem. Consider that the scalar partner of a
right-handed neutrino $N$ possesses a large value during the
inflation. If the value is at the Planck scale and its decay rate is
small, the scalar $N$ dominates the universe before its decay.  Thus,
the decay of the scalar $N$ can produce entropy and dilute the
abundance of the relic gravitino. The crucial point here is that the
scalar $N$ does not decay into a pair of gravitinos due to the matter
(or lepton-number) parity conservation. In other word, $N$ does not
mix with the SUSY breaking field. Thus, this decay process is free
from the gravitino-overproduction problem. Furthermore, the decay of
the scalar $N$ may generate the baryon asymmetry of the
universe~\cite{Murayama:1993em} through the
leptogenesis~\cite{Fukugita:1990gb}.

Another even manifest solution to the gravitino overproduction problem
is to assume the gravitino mass $m_{3/2} <
O(10)\,$eV~\cite{Viel:2005qj}.  In this case, the produced gravitinos
get into thermal equilibrium due to relatively strong interactions
with the standard-model particles, and such light gravitinos are
cosmologically harmless.

Let us comment on another decay mode induced by non-minimal couplings
between the inflaton $\phi$ and the SUSY breaking field $z$. From
(\ref{eq:g-eff-single}) and (\ref{eq:eff-g-multi}), the gravitino
production rate is proportional to $|g_{\bar \phi z z}|^2$. If
$g_{\bar \phi z z}$ is nonzero, the inflaton $\phi$ can also decay
into the SUSY breaking field $z$~\cite{Endo:2006tf}, and the partial
decay rate is comparable to that into the gravitinos.  As noted in
Ref.~\cite{Endo:2006tf}, thus produced $z$ may cause a cosmological
problem at most as severe as that induced by the gravitinos. Therefore
including the effect of the $z$ production may make the problem only a
few times worse, and our discussion remains qualitatively unchanged.

\begin{figure}[t]
\begin{center}
\includegraphics[width=10cm]{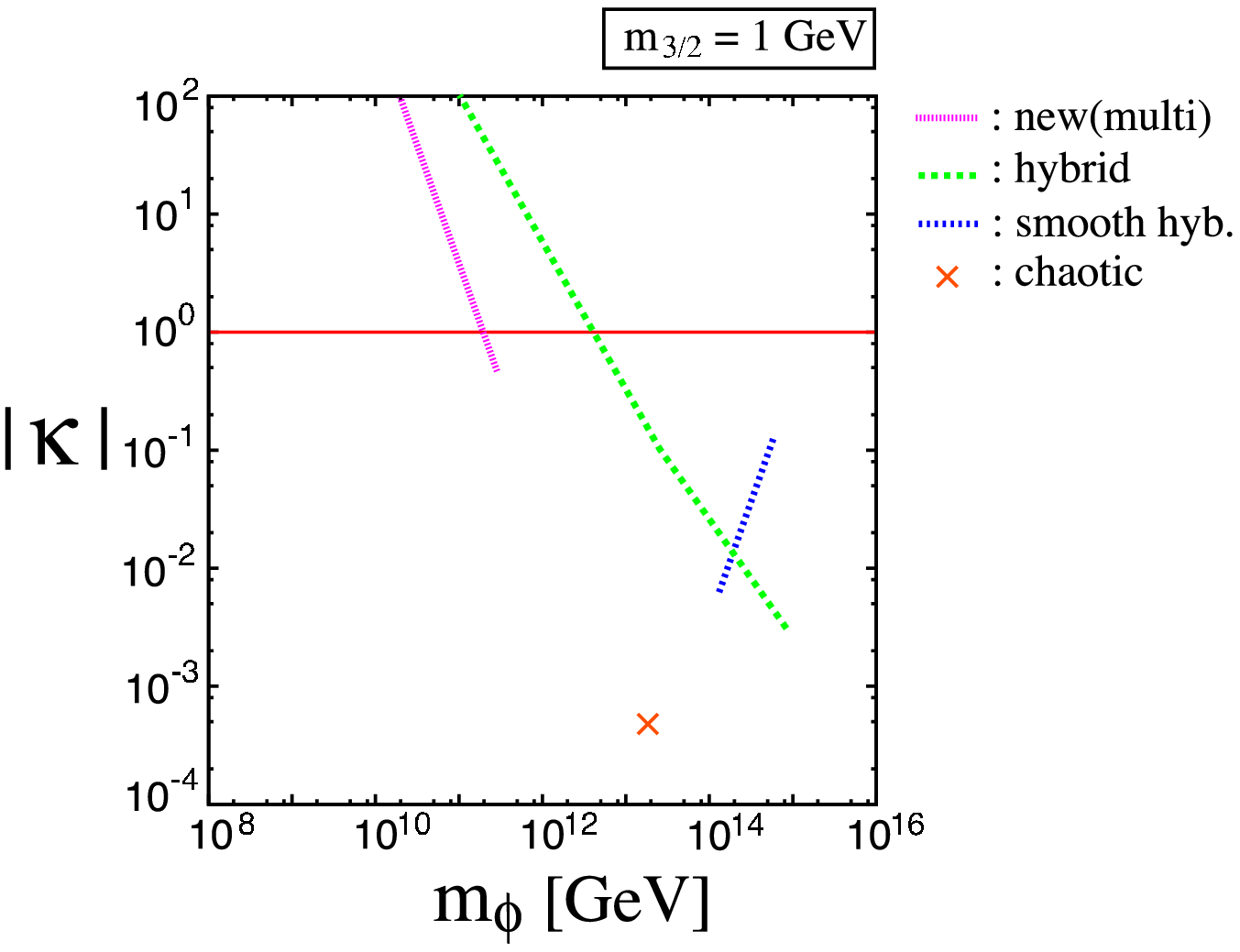}
\caption{Upperbounds on $\kappa$ as a function of the inflaton mass $m_\phi$, with $m_{3/2} =
1{\rm\, GeV}$, for the multi-field new, hybrid, and smooth hybrid, and chaotic inflation
models are also shown.  See the text for the definition of $\kappa$
in each model.
}
\label{kappa1}
\end{center}
\end{figure}

\begin{figure}[t]
\begin{center}
\includegraphics[width=10cm]{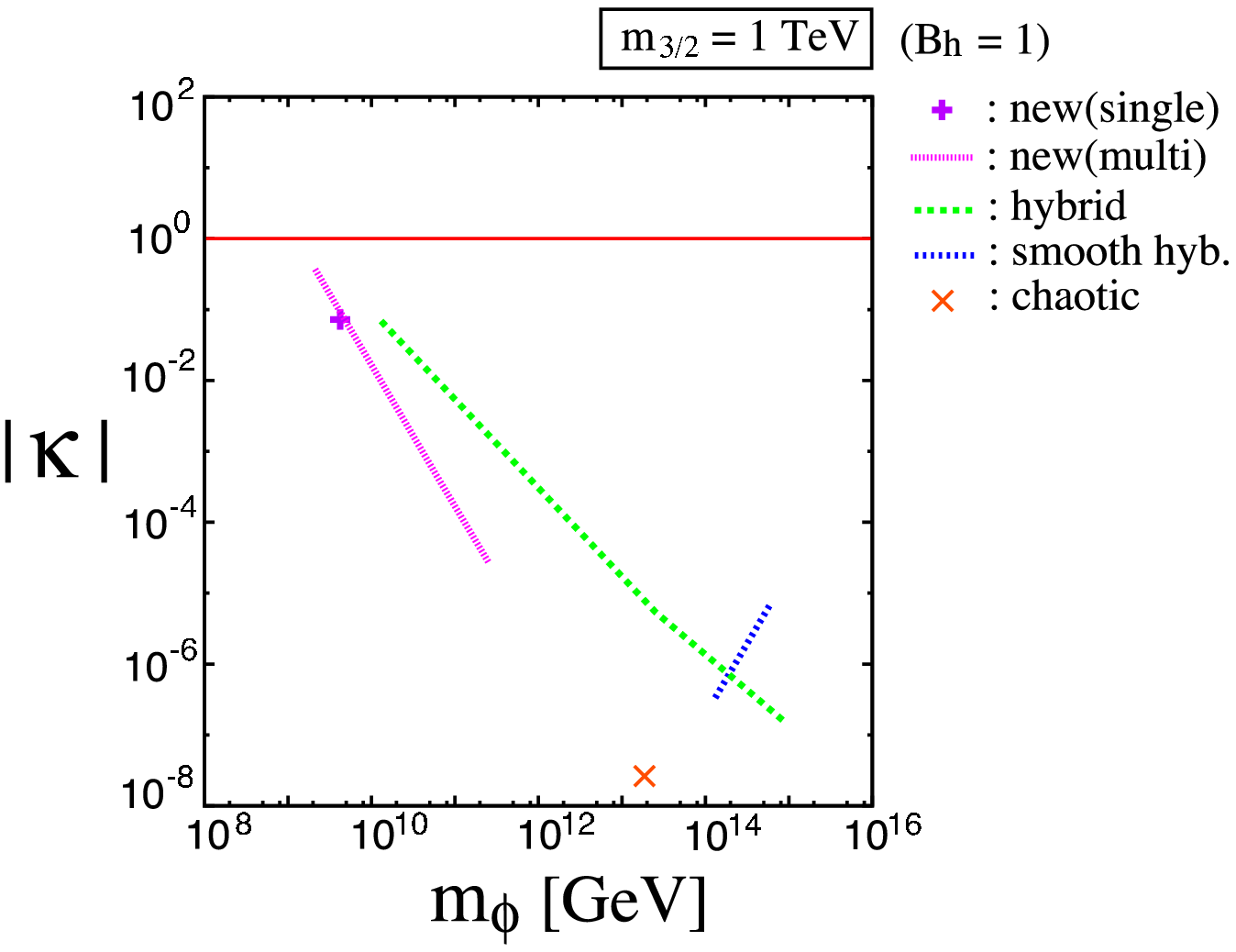}
\includegraphics[width=10cm]{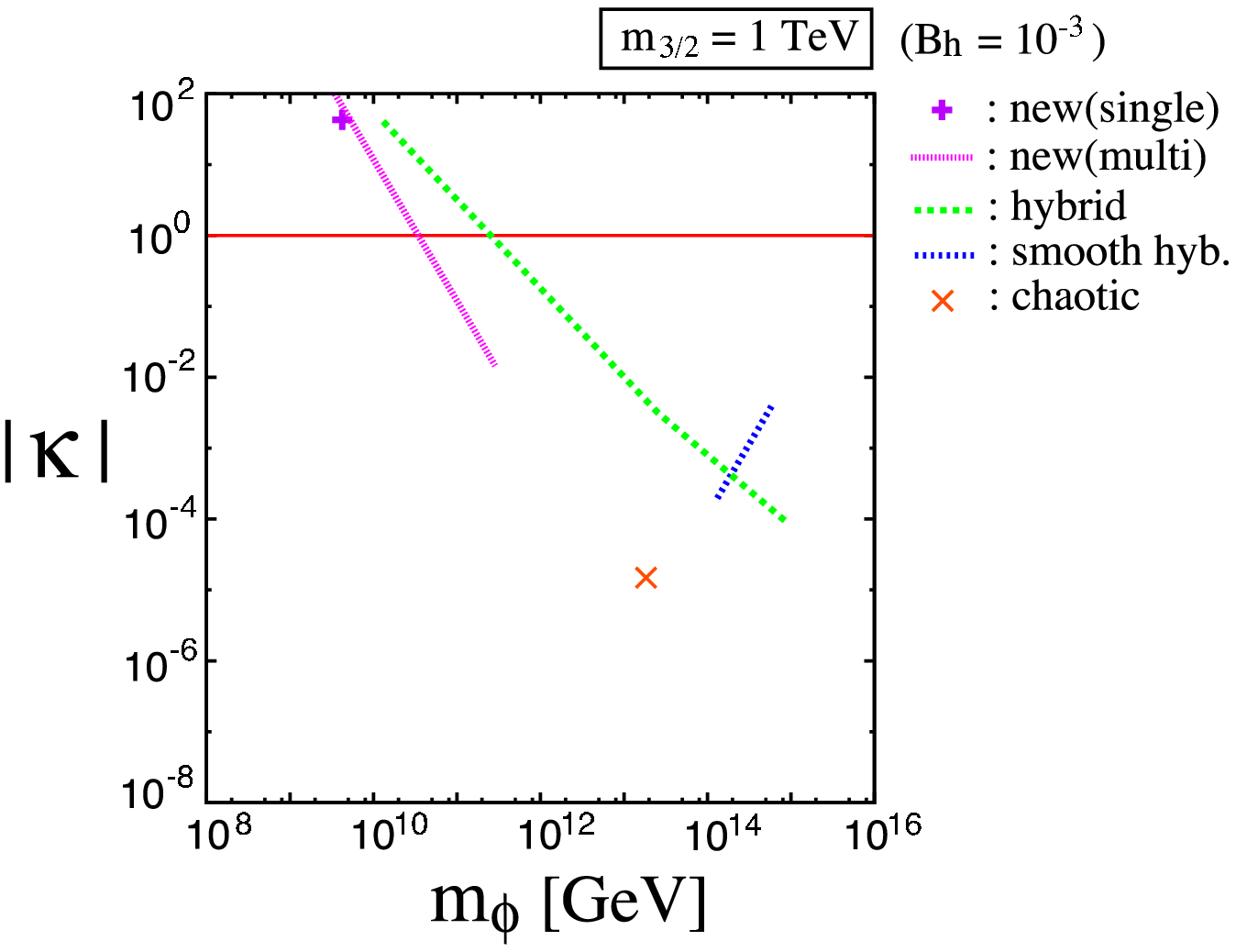}
\caption{Same as Fig.~\ref{kappa1} except for $m_{3/2} =
1{\rm\, TeV}$. The bound on the single-field new inflation model is also plotted.
}
\label{kappa2}
\end{center}
\end{figure}

\begin{figure}[t]
\begin{center}
\includegraphics[width=10cm]{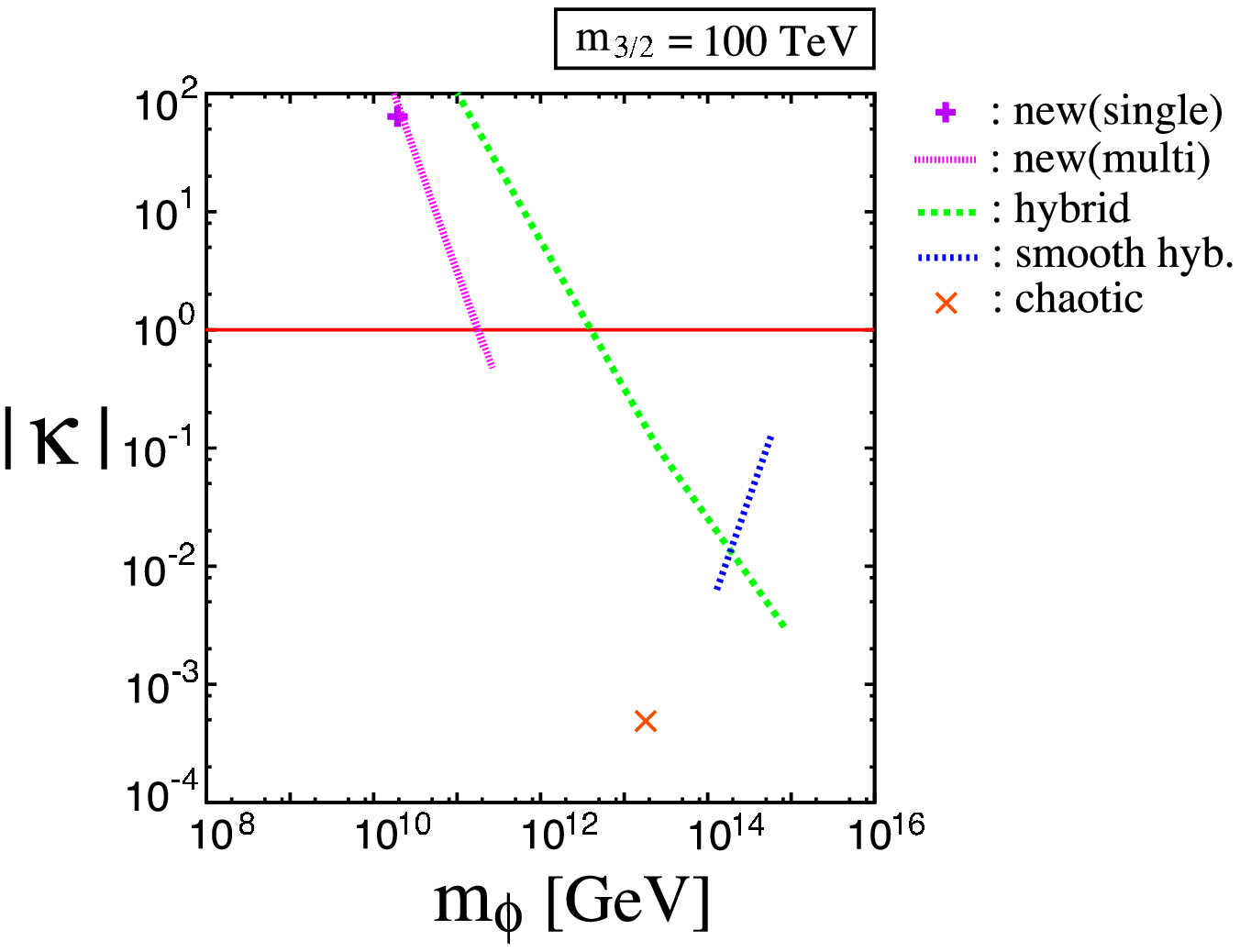}
\caption{Same as Fig.~\ref{kappa2} except for $m_{3/2} = 100{\rm\, TeV}$.
}
\label{kappa4}
\end{center}
\end{figure}

In this paper we have shown that an inflation model generically leads
to the gravitino overproduction, which can jeopardize the successful
standard cosmology. We have explicitly calculated the effective
auxiliary field $\gef$, which is an important parameter to determine
the gravitino abundance, for several inflation models.  The new
inflation is on the verge of being excluded, while the (smooth) hybrid
inflation model is excluded if $\kappa = O(1)$.  To put it
differently, the coefficient of the non-minimal coupling in the
K\"ahler potential, $\kappa$, must be suppressed especially in
(smooth) the hybrid inflation model.  We show the constraints on
$\kappa$ for the inflation model we studied so far in
Figs.~\ref{kappa1} - \ref{kappa4}. As long as the SUSY breaking field
$z$ is singlet, there is no reason that $\kappa$ should be
suppressed. Therefore those inflation models required to have $\kappa
\ll 1$ involve severe fine-tunings on the non-renormalizable
interactions with the SUSY breaking field, which makes either the
inflation models or the SUSY breaking models containing the singlet
$z$ (with $G_z = O(1)$) strongly disfavored.  We stress again that the
existence of such a singlet field is required in the gravity-mediated
SUSY breaking, in order to give the SM gauginos a mass comparable to
the squark and slepton masses.  One of the most attractive ways to get
around this new gravitino problem is to postulate a symmetry of the
inflaton, which is preserved at the vacuum, to forbid the mixing with
the SUSY breaking field.  Among the known models, such a chaotic
inflation model can avoid the potential gravitino overproduction
problem by assuming $Z_2$ symmetry. Another is to assign some symmetry
on the SUSY breaking field $z$ as in the gauge-mediated~\cite{GMSB}
and anomaly-mediated~\cite{Randall:1998uk} SUSY breaking models. So
far we have assumed that $z$ is singlet under any symmetries as in the
gravity-mediated SUSY breaking models.  If the SUSY breaking field $z$
is not a singlet, and the non-minimal coupling like $\delta K =
\kappa/2 |\phi|^2 z z + {\rm h.c.}$ can be suppressed.  It should be
noted however that the mixing between $\phi$ and $z$ may induce other
cosmological problems~\cite{Endo:2006tf} even if $z$ is charged under
some symmetry and/or its VEV is suppressed.

Although we have briefly discussed various (typical) inflation models,
it should be stressed that the gravitino-overproduction problem is
common to all the inflation models in SUGRA.  Thus, in inflation model
building, one must always check whether an inflation model under
consideration satisfies the bound.

\section*{Acknowledgments}
F.T. is grateful to Motoi Endo and Koichi Hamaguchi for a fruitful discussion,
and thanks Q. Shafi for useful communication on the hybrid inflation model.
T.T.Y. thanks M. Ibe and Y. Shinbara for a useful discussion.
F.T.  would like to thank the Japan Society for Promotion of 
Science for financial support.
The work of T.T.Y. has been supported in part by a Humboldt Research Award.


\clearpage


\end{document}